\documentclass[a4paper,11pt]{article}
\pdfoutput=1
\usepackage{jheppub} 
\usepackage{lineno}
\usepackage{bm}\usepackage{physics}
\usepackage{tensor}
\usepackage{autobreak}
\allowdisplaybreaks
\usepackage{tikz}
\usetikzlibrary{
    arrows.meta,
    positioning,
    decorations.markings
}
\tikzset{
    fTSnode/.style={
        rectangle,
        draw=black!60,
        fill=black!5,
        very thick,
        minimum size=5mm},
    fnode/.style={
        rectangle,
        draw=green!60,
        fill=green!5,
        very thick,
        minimum size=5mm},
    TSnode/.style={
        rectangle,
        draw=red!60,
        fill=red!5,
        very thick,
        minimum size=5mm},
    ->-/.style={decoration={
        markings,
        mark=at position .4 with {\arrow{>}}},
        postaction={decorate},
        shorten >=1pt,
        >={Stealth[round]},
        thick},
    -<-/.style={decoration={
        markings,
        mark=at position .6 with {\arrow{<}}},
        postaction={decorate},
        shorten >=1pt,
        >={Stealth[round]},
        thick}
}
\renewcommand\a{\alpha}
\renewcommand\b{\beta}
\renewcommand\d{\delta}

\renewcommand\l{\lambda}
\renewcommand\r{\rho}

\newcommand\e{\epsilon}
\newcommand\g{\gamma}
\newcommand\z{\zeta}
\newcommand\m{\mu}
\newcommand\n{\nu}
\newcommand\x{\xi}
\newcommand\p{\pi}
\newcommand\h{\theta}
\newcommand\s{\sigma}
\newcommand\f{\phi}
\newcommand\w{\eta}

\renewcommand\L{\Lambda}

\renewcommand\S{\Sigma}
\renewcommand\O{\Omega}
\renewcommand\H{\Theta}
\newcommand\D{\Delta}

\newcommand{\fig}[1]{figure~\ref{#1}}
\newcommand{\eq}[1]{eq.~(\ref{#1})}

\newcommand{\eqs}[2]{eqs.~(\ref{#1})-(\ref{#2})}

\newcommand\lb{\left(}
\newcommand\rb{\right)}
\newcommand\ls{\left[}
\newcommand\rs{\right]}
\newcommand\lc{\left\{}
\newcommand\rc{\right\}}
\newcommand{\lan}{\left<}
\newcommand{\ran}{\right>}
\newcommand{\dg}{\dagger}

\newcommand{\non}{\nonumber\\}
\newcommand\pt{\partial}
\newcommand{\idp}[2]{\frac{d^{\,#1}{#2}}{(2\p)^#1}}
\newcommand{\ie}{{i.e.}}
\newcommand{\eg}{{e.g.}}

\newcommand{\cL}{{\cal L}}
\newcommand{\cpt}{{\cal PT}}
\newcommand{\fT}{{\mathfrak{T}}}
\newcommand{\diag}{{\rm{diag}}}
\newcommand{\LE}{{\rm LE}}
\newcommand{\FO}{{\rm FO}}
\newcommand{\gk}{{\g_k}}
\newcommand{\opt}[1]{\widehat{#1}}
\newcommand{\htt}{{\hat t}}
\newcommand{\bx}{{\vec x}}
\newcommand{\bp}{{\vec p}}
\newcommand{\bk}{{\vec k}}
\newcommand{\bq}{{\vec q}}
\newcommand{\bv}{{\vec v}}
\newcommand{\be}{{\vec e}}
\newcommand{\by}{{\vec y}}
\newcommand{\na}{{\nabla}}
\renewcommand{\vec}{\boldsymbol}

\arxivnumber{2412.19416} 

\title{\boldmath Vector and tensor spin polarization for vector Bosons at local equilibrium}

\author[a]{Zhong-Hua Zhang}
\emailAdd{zhzhang22@m.fudan.edu.cn}

\author[a,b,c]{Xu-Guang Huang}
\emailAdd{huangxuguang@fudan.edu.cn}

\author[d]{Francesco Becattini}
\emailAdd{becattini@fi.infn.it}

\author[d]{Xin-Li Sheng}
\emailAdd{sheng@fi.infn.it}

\affiliation[a]{Physics Department and Center for Field Theory and Particle Physics, Fudan University,\\
Songhu Road 2005, Shanghai 200438, China}
\affiliation[b]{Key Laboratory of Nuclear Physics and Ion-beam Application (MOE), Fudan University,\\
Handan Road 220, Shanghai 200433, China}
\affiliation[c]{Shanghai Research Center for Theoretical Nuclear Physics,\\
National Natural Science Foundation of China and Fudan University,\\
Songhu Road 2005, Shanghai 200438, China}
\affiliation[d]{Universit\`a di Firenze and INFN Sezione di Firenze,\\
Via G. Sansone 1, I-50019 Sesto Fiorentino (Florence), Italy}

\abstract{
We derive expressions for the vector and tensor components of the spin polarization of 
massive vector bosons at local thermodynamic equilibrium up to second order in the space-time
gradients of the thermodynamic fields pertaining to the canonical stress-energy tensor and spin
tensor of the free Proca field. A set of Feynman rules is devised to calculate the Wigner 
function and the matrix-valued spin-dependent distribution (MVSD) functions order by order in space-time gradients. Due to constraints imposed by time-reversal symmetry, the leading contribution to spin alignment - 
defined as the 00-component of the tensor polarization - arises from second-order terms in 
MVSD, for which we provide an analytic formula. We discuss the physical meaning of different 
contributions to vector and tensor polarization. These formulae provide a prediction of a 
contribution to the spin alignment which can be compared with the observations in relativistic 
heavy-ion collisions.
}

\begin{document}
\maketitle
\flushbottom

\section{Introduction}
\label{sec:intro}

Spin polarization is a ubiquitous topic across several fields in physics. In atomic 
physics and condensed matter physics, spin has been a frontier research subject for a long time. 
In cosmology, the B-mode polarization of the cosmic microwave background photons can be 
used to test the existence of primordial gravitational waves~\cite{BICEP2:2014owc,BICEP:2021xfz}, 
as well as to elucidate the nature of dark matter and dark energy~\cite{Komatsu:2022nvu}. 
In particle physics, the proton spin problem has triggered extensive discussions that have 
greatly advanced our understanding of Quantum Chromo-Dynamics~\cite{Aidala:2012mv,Ji:2020ena}.

Lately, spin has become an important observable in Quark-Gluon-Plasma physics. Following
early predictions based on partonic angular momentum transfer in relativistic heavy-ion collisions
(HICs)~\cite{Liang:2004ph,Voloshin:2004ha,Gao:2007bc,Huang:2011ru} and on the quantitative
predictions based on statistical local equilibrium~\cite{Becattini:2013fla,Fang:2016vpj,Liu:2020flb}, 
the experiment of the STAR collaboration found evidence of global spin polarization of $\L$ (and $\bar{\L}$)~\cite{STAR:2017ckg,STAR:2018gyt,STAR:2021beb}, 
$\Xi^-$, and $\O^-$ hyperons~\cite{STAR:2020xbm} supporting the predictions based on 
hydrodynamic and statistical local equilibrium. However, these predictions were
not confirmed in the study of local polarization, that is spin polarization as a function of momentum~\cite{Becattini:2017gcx,Wei:2018zfb,Florkowski:2019voj,Wu:2019eyi, Xie:2019jun, Xia:2019fjf,Becattini:2019ntv,Liu:2019krs,Fu:2020oxj,STAR:2019erd,ALICE:2021pzu}. It was later 
found out that further terms which are linear in the gradients of the thermo-hydrodynamic fields
are responsible for spin polarization at local equilibrium (spin-thermal shear and spin 
Hall effect) which can reproduce the local polarization data;
for recent reviews, see~\cite{Liu:2020ymh,Becattini:2020ngo,Gao:2020vbh,Huang:2020dtn,Jian-Hua:2023cna,Niida:2024ntm,Becattini:2024uha}. 

Similar spin polarization phenomena could occur in HICs for vector mesons~\cite{Liang:2004xn}. 
In this case, the benchmark quantity is the so-called spin alignment parameter, $\H_{00}-1/3$. 
Here, $\H_{00}$ is the $00$-element of the spin density matrix of a vector meson, representing 
the probability of the vector meson being in the state with spin-0 in its rest frame in a specified quantization direction, \ie the longitudinally polarized mode (see section~\ref{sec:massive:spin} 
for discussions of the spin state of vector bosons). Spin alignment is essentially a component of vector meson's tensor polarization, which will be discussed in detail in section~\ref{sec:massive:spin}.

Recently, the STAR Collaboration reported the measurements of the global spin alignment of 
$ \phi(1020) $ and $ K^{*0}(892) $ 
mesons~\cite{STAR:2022fan}, \ie, the average spin alignment with respect to the direction of the
global orbital angular momentum (OAM) of the colliding system (which coincides with the direction 
of the reaction plane for symmetric collisions such as Au + Au). They found that 
$ \H_{00} - 1/3 >0 $ for $ \phi $ meson and $ \H_{00} - 1/3 \approx 0 $ for $ K^{*0} $ meson 
at BNL Relativistic Heavy Ion Collider (RHIC) energies. On the other hand, the ALICE Collaboration reported that $ \H_{00} - 1/3<0 $  
for both  $ \phi $ and $ K^{*0} $ in the low-$p_T$ region at the LHC energy~\cite{ALICE:2019aid}. 
A substantial spin alignment of $J/\psi(3097)$ meson with $ \H_{00} - 1/3<0 $ at low-$p_T$ 
has also been observed by the ALICE Collaboration~\cite{ALICE:2020iev,ALICE:2022dyy}. 

Spin alignment is a challenging puzzle in HICs because the results depend strongly on both collision energies and meson type~\cite{STAR:2022fan}.
Moreover, its magnitude for $\phi$ meson is large ($\mathcal{O}(10^{-2})$), which is at variance with extrapolations from the
hydrodynamic-local equilibrium model and other models proposed in the literature.
For instance, in a quark coalescence model, supposing that quarks coalesce to form mesons and that their spins are at local equilibrium at the time of hadronization, one may naively expect that the spin alignment of vector mesons would be of the order of $P_q^2$ with $P_q$ the spin polarization of quarks and antiquarks. 
However, this is of the order of $\mathcal{O}(10^{-4})$ assuming that $P_q$ is about the same 
magnitude of the global spin polarization of $\Lambda$~\cite{STAR:2017ckg}. In view of these failures,
theorists have proposed various mechanisms beyond local equilibrium calculations, such as contributions 
from electromagnetic field~\cite{Yang:2017sdk,Sheng:2019kmk,Xia:2020tyd,Sheng:2022ssp,Li:2023tsf,Zhao:2024ipr}, strong $\phi$ mean field~\cite{Sheng:2019kmk,Sheng:2022wsy,Sheng:2023urn}, medium modification to meson spectra~\cite{Li:2022vmb,Wei:2023pdf,Fu:2023qht,Sun:2024anu,Li:2024qae}, and gluonic color fields~\cite{Kumar:2022ylt,Kumar:2023ghs}, based on various theoretical approaches such as quantum kinetic theories~\cite{Yang:2021fea,Sheng:2022ffb,Wagner:2022gza,Kumar:2023ghs,Kumar:2023ojl,Yin:2024dnu}, effective models~\cite{Sheng:2022ssp,Wei:2023pdf,Sun:2024anu}, linear response theories~\cite{Li:2022vmb,Dong:2023cng}, and holographic models~\cite{Zhao:2024ipr,Sheng:2024kgg}. 

In a quantum hydrodynamic picture, a full calculation of spin alignment
has not been carried out yet. An estimate in ref.~\cite{Sheng:2019kmk} shows that the contribution to $ \H_{00} -1/3 $ from B-part (E-part) of fluid vorticity is around $ \mathcal{O}(10^{-5}) $ ($\mathcal{O}(10^{-4})$). But it was based on a quark coalescence scenario with quarks polarized by the vorticity field without incorporating the direct coupling between meson's spin and vorticity. On the other hand, most of the aforementioned hydrodynamic estimates rely on formulae 
that apply to global equilibrium, where the thermal shear tensor vanishes. At local thermodynamic 
equilibrium, the contribution from thermal shear tensor to vector meson spin alignment, unlike to fermion spin polarization~\cite{Becattini:2021suc,Liu:2021uhn}, also vanishes at the linear order, as will be shown later. Nevertheless, there might be a non-vanishing dissipative contribution at the linear order in the thermal
shear; a particular calculation has been carried out in ref.~\cite{Li:2022vmb} which, however, shows 
a critical dependence on the lifetime of the vector meson. This was confirmed in the framework of relativistic
kinetic theory \cite{Wagner:2022gza,Dong:2023cng}.

In this article, we use quantum statistical methods, notably the Local Equilibrium Density Operator (LEDO) \cite{Zubarev:1979,Weert:1982,Becattini:2019dxo}, to derive a formula (``Cooper-Frye type") for the vector and
tensor spin polarization of vector bosons at local equilibrium, which is complete up to the second order
in space-time gradients. Quantum statistical methods take into account the effects of non-constant
thermodynamic fields in a very effective way and, particularly, they show that even local equilibrium 
involves gradient corrections to the distribution functions. This method has been used to derive the 
spin-thermal shear coupling~\cite{Becattini:2021suc,Buzzegoli:2022fxu} and the mean spin  
polarization vector for spin-1/2 fermions at both first order~\cite{Liu:2021nyg,Buzzegoli:2021wlg} and 
second order~\cite{Sheng:2024pbw} in space-time gradients. The second-order gradient corrections we report 
in this work most likely include those found in ref.~\cite{Kumar:2023ojl} in that they are obtained by expanding 
the local equilibrium density operator.
We note that, even though we identify gradients of the thermal vorticity or thermal shear tensors $\partial\varpi$ and $\partial\xi$ as the second-order terms in our power counting scheme, their magnitude could be large due to significant spatial variation of $\varpi$ and $\xi$ around the freeze-out hypersurface. Their contributions to the spin alignment could be important in realistic heavy-ion collisions, which need to be investigated in numerical simulations.

It should be emphasized that, as the calculation is at local 
equilibrium, its result depends on the specific form of the stress-energy tensor and spin tensor, the 
so-called pseudo-gauge choice \cite{Becattini:2018duy}. In this work, we focus on the canonical stress-energy
and spin tensors obtained from the familiar Proca Lagrangian (see \eq{Lag:free}) and we
allow a spin potential $\Omega$ differing from thermal vorticity $\varpi$. It should be pointed out 
that the corresponding expressions in the Belinfante pseudo gauge cannot be obtained from our results 
by simply setting $\Omega=\varpi$.

The paper is organized as follows.
In section~\ref{sec:massive:spin}, we briefly discuss the definition of the Matrix-Valued Spin-dependent Distribution (MVSD) and the description of the spin state of massive bosons using the spin density matrix.
In section~\ref{sec:ledo}, we introduce LEDO and its cumulant expansion based on our power counting 
rules which will be given accordingly.
In section~\ref{sec:MVSDtop}, we show how to extract the vector and tensor polarization from the MVSD.
The zeroth- and first-order results of the MVSD are given in section~\ref{sec:zeroth:dis}.
In section~\ref{sec:diagram}, we introduce a Feynman diagram scheme and apply it to the calculation 
of the second-order terms.
The second-order MVSD is given in section~\ref{sec:2nd:order} with diagrammatic representations.
In section~\ref{sec:check}, we take the global equilibrium limit and obtain the spin alignment at global equilibrium. We also derive a new contribution from the space-time gradients of the thermal vorticity beyond global equilibrium.
We summarize our results and give an outlook in section~\ref{sec:summary}.

Throughout this paper, we use the Minkowski metric $\w_{\m\n}=\diag(1, -1, -1, -1)$ and adopt natural units $ k_B = c = \hbar =1 $. The Levi-Civita tensor $\e^{\m\n\r\s}$ is normalized as $\e^{0123}=-\e_{0123}=1$. A bigger hat over a symbol, such as $ \opt{O} $, denotes an operator in Hilbert space. A smaller hat, such as $ \hat{k} $, denotes a suitably normalized four-vector, for example, $ \hat{k}^\m = k^\m / (\htt\cdot k) $ with $ \htt^\m = (1,{\bm 0}) $ being the normal vector in the time direction. We also define the shorthand notations $ A^{[\m}B^{\n]} = A^{\m}B^{\n}-A^{\n}B^{\m} $  and  $ A^{(\m}B^{\n)} = A^{\m}B^{\n}+A^{\n}B^{\m} $ for arbitrary two vectors $A^\mu$ and $B^\mu$.

\section{Spin state of a massive vector boson}
\label{sec:massive:spin}
We consider free neutral massive vector bosons, such as $\f$  and $J/\psi$, which are described by the Proca Lagrangian
\begin{equation}
    \label{Lag:free}
    \cL = -\frac{1}{4} F_{\m\n}F^{\m\n}+\frac{1}{2}m^2A_\m A^\m
\end{equation}
with $m$ being the boson's mass and the field tensor
\[
   F^{\mu\nu} \equiv \pt^\mu A^\nu - \pt^\nu A^\mu.
\]
The equations of motion (EOMs) for massive vector field $A^\m$ obtained from \eq{Lag:free} are given by
\begin{gather}
    \label{eom:massive}
    (\pt^2+m^2)A^\m=0 \,,\\
    \label{eom:lorentz}
    \pt_\m A^\m =0 \,,
\end{gather}
where the second equation is the Lorenz condition for the massive vector field.

The mode decomposition of the vector field $A^\m$ after quantization reads,
\begin{equation}
    \label{proca:field:decomp}
    \opt{A}^\m(x) = \sum_{s} \int\frac{d^3\bk}{(2\p)^3}\frac{1}{2E_\bk} \ls \opt{a}_\bk^s \e^\m_s(k) e^{-ik\cdot x} +\opt{a}_\bk^{s\dag} \e_s^{\m*}(k) e^{ik\cdot x}\rs,
\end{equation}
where $ k^0 = E_\bk = \sqrt{\bk^2+m^2} $, $ \e_s^\m(p) $ is the polarization vectors~\footnote{Note that $ \e_s^\m(p) $ depends only on $\bp$, but for later use, it is convenient to consider $E_\bp$ in $\e_s^\m(p)$ as $p^0$.}, and $ \opt{a}_\bk^s $ and $ \opt{a}_\bk^{s\dag} $ are annihilation and creation operators for vector boson of momentum $\bk$ and spin state $s$. They satisfy the commutation relation
\begin{equation}
    \label{photon:commutation}
    [\opt{a}_\bk^s, \opt{a}_{\bk'}^{s'\dag}] = (2E_\bk) (2\p)^3 \d^{(3)}(\bk-\bk') \d^{ss'}.
\end{equation}
The decomposition in \eq{proca:field:decomp} can be straightforwardly extended to the case of 
charged vector bosons (such as $K^{*0}$ and $D^{*+}$ mesons) by introducing different operators 
for particles and antiparticles.

The physical meaning of $s$, which is a quantum number taking values $-1,0,1$ depends on the choice 
of the so-called {\em standard Lorentz transformation} $[k]$, which is defined as the Lorentz transformation
turning the standard vector $k_0^\mu = (m,{\bf 0})$ of a particle with mass $m$ into its actual 
four-momentum $k$. As it is well known, if $[k] = R_z(\phi) R_y(\theta) L_z(\xi)$ where $\phi,\theta,\xi$
are the polar and hyperbolic coordinates of the four-vector $p$, $R_i$ are rotations around the
axis $i=x,y,z$ and $L_i$ boosts along axis $i$, $s$ is the helicity of the particle. 
On the other hand, if $[k]$ is chosen to be a pure Lorentz boost, $s$ can be thought of as a spin
component along the boosted quantization axis. It is important to stress that both the creation and
annihilation operators as well as the polarization vectors $\epsilon_s$ depend on the choice of the standard Lorentz transformation,
so that, strictly, one should write
\[
\epsilon^\mu_s([k]) \qquad \qquad \widehat{a}^s([k])
\]
even though their combination, summed over $s$, in the field expansion is altogether independent 
thereof. Throughout this paper, this dependence will not be explicitly shown in the notation, and
we will keep indicating $\opt{a}_\bk^s$ and $\e^\m_{s}(k)$. Of course, if $ k^\m = k^\m_0 $, there is no 
dependence on the standard Lorentz transformation, which is the identity in all cases.

The Lorentz condition in \eq{eom:lorentz} enforces the polarization four-vectors to satisfy 
$k_\m \e_s^\m(k)=0$, and such a condition ensures that $\e^\m_s(k)$ in boson's rest frame has 
a vanishing time component. Therefore, there are only three linearly independent $\e^\m_s(k_0)$ 
vectors, which can be chosen as the basis vectors of the spin-1 representation space,
with, \eg, the $y$ axis as the spin quantization direction. In this case, these vectors are the
eigenvectors of the representation matrix of the generator $J_y$, that is
\[
\begin{cases}
    \displaystyle\be_{1} =-\frac{1}{\sqrt{2}}(i,0,1) \,,\\
    \displaystyle\be_{0} =(0,1,0) \,,\\
    \displaystyle\be_{-1} =\frac{1}{\sqrt{2}}(-i,0,1) \,.
\end{cases}
\]
The polarization four-vectors are then $\e^\mu_s(k_0) =(0,\be_s) $ and the $\e^\m_s(k)$ can be 
obtained by applying the standard Lorentz transformation $[k]$, that is $\e^\mu_s(k) = 
[k]\indices{^\mu_\nu} \e^\nu_s(k_0) $.
Furthermore, one can also introduce the ``scalar" polarization vector as $\e^\m_o(k_0)=\htt^{\m}$  
in the rest frame, with $ \htt^\m =(1,\bm 0)$.
Thereby, the four four-vectors $\e^\m_1$, $\e^\m_0$, $\e^\m_{-1}$, $\e^\m_o$ at either $k_0$ or $k$
form a {\em vierbein} that satisfies the following orthonormality and completeness relations
\begin{equation}
    \label{vierbein}
    \w^{ab}\e_a^{\m} \e_b^{\n*}=\w^{\m\n},\quad \w_{\m\n}\e^{\m}_a \e^{\n*}_b=\w_{ab} \,,
\end{equation}
where $a, b$ take the values from $o,0,\pm1$, $ \w^{ab} $ and $ \w_{ab} $ are defined
as
\begin{equation}
    \w_{ab}=\w^{ab}=
    \begin{cases}
        1 & a=b=o,\\
        -1 & a=b=0,\pm1 \\
        0 & \text{else}.
    \end{cases}
\end{equation}
For a reason that will become clear below, it is convenient to choose $[k]$ as a pure boost $L(k)$
carrying the four-momentum $k^\mu_0$ to $k^\mu$, so that $\e_s^\m(k) =L\indices{^\m_\n}(k) \e^\n_s(k_0) $
with
\begin{equation}\label{eq:Lorentz-transform}
    L^0_{~0}(k)=E_\bk/m\,, \qquad L^0_{~i}(k)=L^i_{~0}(k)=k^i/m\,, \qquad L^i_{~j}(k)=\d^i_j-k^i k_j/[m(E_\bk+m)]\,.
\end{equation}
Therefore,
\begin{equation}\label{polar:vectors}
    \e_s^\m(k)=\lb \frac{\bk\cdot{\bm e}_s}{m} \,,\, {\bm e}_s + \frac{\bk\cdot{\bm e}_s}{m(E_\bk+m)}\bk\rb \,,
    \quad s=\pm1,0 \,.
\end{equation}
or, in a manifestly covariant form
\begin{equation}\label{polar:conva}
    \e_s^\m(k) = \e^\m_s(k_0) -\frac{k\cdot \e_{s}(k_0)}{m(\htt\cdot k+m)}(k^\m+m \htt^\m) \,,
    \quad s = \pm1,0 \,.
\end{equation}
Since the $\e_s^\m(k)$ are perpendicular to the four-vector $k^\m$, the polarization vectors  
$\e_s^\m(k),\ s=0,\pm 1$ form a set of a complete basis orthogonal to $k^\mu$ (which can also be obtained from the first relation in \eq{vierbein}),
\begin{equation}
    \sum_{s=-1}^{+1}\e_s^\m(k)\e_s^{\n*}(k) = -\w^{\m\n}+\frac{k^\m k^\n}{m^2} \,.
\end{equation}

The spin polarization state of a relativistic particle is described by the so-called
spin density matrix. In a relativistic quantum field theory, this matrix can be defined
as \cite{Becattini:2020sww}
\begin{equation} \label{Spin-density-matrix}
    \H_{rs}(k) = \frac{\Tr \lb \opt{\r}\,\opt{a}^{s\dg}_{\bk} \opt{a}^{r}_{\bk}\rb}
    {\sum_r \Tr \lb \opt{\r}\, \opt{a}^{r\dg}_{\bk} \opt{a}^{r}_{\bk} \rb} \,,
\end{equation}
where $r,s$ are the spin-state indices running from $ -S $ to $ S $ with $S$ the particle spin and 
$k$ is the four-momentum of the particle in a conventional global frame (\eg~the laboratory frame). 
The spin density matrix can be used to express all measurable quantities related to the spin of 
the particle. However, the spin density matrix, featuring a quadratic combination of creation
and annihilation operators, also depends on the choice of the standard Lorentz transformation. 
Notably, it can be shown (see appendix~\ref{sec:STDL}) that if one defined it by using a
different standard Lorentz transformation $[k]'$, then the new spin density matrix would be
\[
\Theta^\prime(k)_{tu} = \sum_{rs} D^S(R^{-1})_{tr} \Theta(k)_{rs} D^S(R)_{su} \,,
\]
where $R \equiv [k]^{-1}[k]'$ is a rotation as it keeps the time direction unchanged,
and $D^S$ stands for the spin-$S$ irreducible representation of the SO(3) group.

Even though the spin density matrix depends on the standard Lorentz transformation, measurable
physical quantities must be independent thereof. For instance, it can be shown that the spin 
polarization four-vector in the conventional frame (\ie~the frame where 
the four-momentum of the particle is $k$), defined as the mean value of the Pauli-Lubanski
vector reads \cite{Becattini:2020sww}
\begin{equation}\label{spinpolvec}
S^\mu(k) = \sum_{i=1}^3 \tr \left( D^S(J^i) \Theta(k) \right) [k]\indices{^\mu_i} \,,
\end{equation}
where $J^i$ are the generators of the rotation group (also known as angular momenta) 
and $D^S(J^i)$ their associated matrices in the representation of spin-$S$. The four-vector $ S^\mu(k) $ in
\eq{spinpolvec} turns out to be independent of the choice of $[k]$ (see appendix~\ref{sec:STDL}),
hence $ S^\mu(k) $ is that of an actual physical vector.
To determine the spin polarization vector in the rest frame of the particle, which
is the measurable quantity, a Lorentz transformation is to be carried out. This 
is usually done keeping in the rest frame the same space axes of the conventional frame: 
otherwise stated with a pure Lorentz boost $L(k)^{-1}$. Looking at \eq{spinpolvec},
it can be then concluded that the spin polarization vector in the rest frame can be
expressed as
\begin{equation}\label{spinpolvec2}
    S^i_\text{rest}(k) = \tr \left( D^S(J^i) \Theta(k) \right)\,,
\end{equation}
(with the time component vanishes) only if the standard Lorentz transformation $[k]$ in
\eq{spinpolvec} is a pure boost $L(k)$. Note that in \eq{spinpolvec2}, the
spin vector components are in the rest frame of the particle, but its argument $k$ is the
four-momentum in the conventional observer frame.

Similar arguments apply to all other physical quantities that can be formed out of the 
spin density matrix, like tensors. For the spin-1 case, being $\H(k)$ a $ 3\times 3 $ Hermitian 
matrix with unitary trace, it can be decomposed as follows
\begin{equation}
    \label{dmatrix:decomp}
    \H(k)=\frac{1}{3} I+\frac{1}{2}\sum_{i=1}^3 S^i_\text{rest}(k) S^i + \frac{1}{4}
     \sum_{i,j=1}^3 \fT^{ij}(k) \S^{ij} \,,
\end{equation}
where $S^i \equiv D^1(J^i)$ (fulfilling $[S^i, S^j]=i \sum_k \e^{ijk}S^k$), $ \vec{S}^2 = 2 I $ and
$\S^{ij} = S^i S^j +S^j S^i -\frac{2}{3} \vec{S}^2 \delta^{ij}$.
Choosing $\hat{\bm y}$ as the spin quantization axis, one has
\begin{equation}
    \label{eq:si}
    S^1=S_{x} =\frac{i}{\sqrt{2}}
    \begin{pmatrix}
        0 & -1 & 0\\
        1 & 0 & -1\\
        0 & 1 & 0
    \end{pmatrix},\quad
    S^2=S_{y} =
    \begin{pmatrix}
        1 & 0 & 0\\
        0 & 0 & 0\\
        0 & 0 & -1
    \end{pmatrix},\quad
    S^3=S_{z} =\frac{1}{\sqrt{2}}
    \begin{pmatrix}
        0 & 1 & 0\\
        1 & 0 & 1\\
        0 & 1 & 0
    \end{pmatrix} \,.
\end{equation}
Since
\begin{align*}
 &   \tr S^i = \tr \S^{ij} = 0 \qquad \qquad  \tr (S^i S^j) = 2 \delta^{ij}  \\    
 &   \tr (S^i \S^{jk}) = 0 \qquad \qquad
 \tr (\S^{ij} \S^{lk}) = 2 (\delta^{il} \delta^{jk} + \delta^{ik}\delta^{jl} ) \,,
\end{align*}
one can readily invert the above formulae
\begin{align}
    \label{exp:P}
    S^{i}_\text{rest}(k) &= \tr \left( S^i \H(k) \right) \,,\\
    \label{exp:T1}
    \fT^{ij}(k) &= \tr \left( \S^{ij} \H(k) \right) \,.
\end{align}
Therefore, comparing with \eq{spinpolvec2} and recalling the discussion
of its meaning, it turns out that the $S_\text{rest}^i(k)$ in \eq{dmatrix:decomp}
(in accordance with the chosen notation) are just the components of the spin  vector 
in the rest frame of the particle, provided that the standard Lorentz transformation is a pure 
boost. Similarly, the proper physical polarization tensor should be defined in the frame 
where the particle four-momentum is $k$, and it turns out to be
\[
    \fT^{\mu\nu}(k) = \sum_{ij} \tr (\Sigma^{ij} \Theta(k)) ~ [k]\indices{^\mu_i} ~ [k]\indices{^\nu_j} \,,
\]
which is independent of the standard Lorentz transformation.
Equation (\ref{exp:T1}) provides the components of the same tensor in the particle rest frame only if the standard 
Lorentz transformation is a pure Lorentz boost.

Instead of the Cartesian components of the above operators, we can use their spherical irreducible
(under the rotation group SO(3)) components. Denoting $S_{1,m}$ and $ S_{2,m} $ the spherical 
components of $\vec{S}$, that is $ S_{1,0}=S_{y}, S_{1,\pm1}=\mp(S_{z}\pm iS_{x})/\sqrt{2} $ and
\begin{equation}
    \label{def:s2}
    S_{2,m} = \sum_{m_1,m_2=-1}^1 \braket{1,m_1;1,m_2}{2,m}S_{1,m_1}S_{1,m_2} \,,
\end{equation}
where $ \braket{1,m_1;1,m_2}{2,m} $ are the Clebsch-Gordon coefficients; and further denoting
\begin{equation}
    \label{exp:T2}
    P_{1,m}(k)= \tr \lc S_{1,m}\H(k) \rc \qquad \qquad
    \fT_{2,m}(k) = \tr \lc S_{2,m}\H(k) \rc
\end{equation}
the relevant traces of the spin density matrix, one has
\begin{equation}\label{dmatrix:irdecomp}
    \H(k) =\frac{1}{3}I +\frac{1}{2}\sum_{m=-1}^1 (-1)^m P_{1,-m}(k) S_{1,m} 
    +\sum_{m=-2}^{2}(-1)^{m} \fT_{2,-m}(k) S_{2,m}.  \;\; ;
\end{equation}
for more details, see ref.~\cite{Sakurai:2011zz}. After lengthy calculations, one can show that
\[
    \H_{00}-1/3 = -\sqrt{2/3}\fT_{2,0} 
\]
which is, by definition, the \textit{spin alignment} parameter of the vector boson associated 
with a given spin quantization direction (here, $\hat{\by}$).

The irreducible tensor $\fT_{2,m}$ plays a major role in the decay of polarized vector 
mesons. For example, in the decay $ \f\rightarrow K^{+} + K^{-} $, this tensor determines
the angular distribution $(1/N)(dN/d\O)(\h, \f)$ of either daughter particle. 
Due to the transformation property under rotation, the angular distribution must have the following form
\begin{equation}
    \frac{1}{N}\frac{dN}{d\O}(\h, \f) = \frac{1}{4\p} +\a\sum_{m=-1}^{1}(-1)^m
    P_{1,m}Y_{1,-m}(\h,\f) +\b\sum_{m=-2}^{2}(-1)^m\fT_{2,m}Y_{2,-m}(\h,\f) \,,
\end{equation}
where $ (\h, \f) $ are the polar and azimuthal angles with respect to the spin quantization 
axis in the rest frame of the mother particle and $ Y_{l,m}(\h,\f) $ the spherical harmonic 
functions. One can show that $ \a=0 $  for strong decays because of the pairty conservation, 
and $ \b=-\sqrt{\frac{3}{10\p}} $  for pseudoscalar meson productions, which is obtained after 
explicit calculation using the Wigner D-matrix formalism~\cite{Schilling:1969um,Yang:2017sdk}.
Therefore, only the tensor polarization $ \fT_{2,m} $ can be measured in the pseudo-scalar 
meson decay modes of the vector mesons. If we integrate over the azimuthal angle, we have
\begin{equation}
    \frac{1}{N}\frac{dN}{d\cos{\h}} = \frac{1}{4\p} -\frac{1}{4\p}(3\cos^2{\h}-1)\sqrt{\frac{3}{2}}\fT_{2,0}.
\end{equation}
Thus, $ \H_{00}-1/3 = -\sqrt{2/3}\fT_{2,0} $ is the only parameter that determines the polar angle distribution of the decay daughters in the rest frame of the vector meson.

For the purpose of the calculations of polarization in a relativistic fluid, it is very useful 
to introduce the Matrix-Valued Spin-dependent Distribution (MVSD) operator
\begin{equation}
\label{def:MVSD}
    \opt{f}_{rs}(x,k) = \frac{1}{2\p}\e_{\m}^{r*}(k)\e_{\n}^{s}(k) \opt{W}_{+}^{\m\n}(x,k)
\end{equation}
with $ \opt{W}_{+}^{\m\n}(x,k) $ standing for the positive-energy (particle) part of the Wigner operator
\begin{equation}
\label{def:Wig}
    \opt{W}_{+}^{\m\n}(x,k) = \int d^4 y\,e^{ik\cdot y} \opt{A}^{\n}\left( x-\frac{y}{2} \right) \opt{A}^{\m} \left(x+\frac{y}{2} \right) \h(k^2)\h(k^0) \,,
\end{equation}
where $ \h $ is the Heaviside step function.
With \eq{proca:field:decomp}, we obtain
\begin{equation}
\begin{split}\label{Wigner-operator}
    \opt{W}_{+}^{\m\n}(x,k) =&~ \int\idp{3}{\bp_1} \idp{3}{\bp_2}\frac{1}{2E_{\bp_1}}\frac{1}{2E_{\bp_2}}(2\p)^4 \sum_{a_1a_2}e^{-i(p_1-p_2)\cdot x} \\
    &~ \times \d^{4}\lb k-\frac{p_1}{2}-\frac{p_2}{2}\rb \opt{a}^{a_2\dg}_{\bp_2}\opt{a}^{a_1}_{\bp_1}\e^{\m}_{a_1}(p_1)\e^{\n*}_{a_2}(p_2) \,.\\
\end{split}
\end{equation}
In general, the argument $k$ is not on the mass shell in \eq{def:MVSD} and \eq{Wigner-operator}. However, since 
$p_1^\mu$ and $p_2^\mu$ lie on the mass shell, we have $ (p_1-p_2)\cdot(p_1+p_2) =0 $. Then, the 
Wigner operator fulfills the equation $ (k\cdot\pt_x)\opt{W}_{+}^{\m\n}(x,k) =0 $ \cite{Becattini:2020sww}. 
Likewise, from the \eq{def:MVSD}, we have
\begin{equation}\label{kdiv}
    \lb k\cdot\pt_x\rb \opt{f}_{rs}(x,k)=0 \,.
\end{equation}
It can be shown in ref.~\cite{Becattini:2020sww} that this equation makes the argument $k$ an on-shell
four-vector {\em after} integration over an arbitrary 3D hypersurface $\S$ with suitable 
boundary conditions. From eqs.~\eqref{def:MVSD},~\eqref{Wigner-operator}, and~\eqref{Spin-density-matrix} it can be shown that
\[
 \int_\S d \S \cdot k \; \Tr (\opt{\r}\,\opt{f}_{rs}(x,k)) = \frac{1}{2} \d(k^2-m^2) \h(k^0) \Tr\lb \opt{\r}\,\opt{a}^{s\dg}_{\bk}\opt{a}^{r}_{\bk}\rb \,,
\]
so that the spin density matrix in \eq{Spin-density-matrix} can be expressed as 
an integral over an arbitrary 3D hypersurface $\S$
\begin{equation}\label{eq:cooper-frye}
     \H_{rs}(k) = \frac{ \int_\S d \S \cdot k \; \Tr (\opt{\r}\,\opt{f}_{rs}(x,k))}
    {\sum_r \int_\S d \S\cdot k \; \Tr (\opt{\r}\,\opt{f}_{rr}(x,k))} \,.
\end{equation}
\eq{kdiv} ensures that the four-vector $k$ is on-shell, \ie $k^2=m^2$.
\eq{eq:cooper-frye} is a very useful form for the applications to relativistic 
hydrodynamics because the MVSD, unlike the spin density matrix overall, depends on $x$ and its 
value is mostly determined by values of hydrodynamic fields around $x$ at local thermodynamic equilibrium.

It is also possible to parametrize the MVSD as the product
\begin{equation}\label{fdecomp}
    f_{rs}(x,k)=\H_{rs}(x,k) f(x,k) \,,
\end{equation}
where $ f(x,k) = \sum_r f_{rr}(x,k) $ is the trace of the MSVD, so that $\tr \H(x,k)=1$.
The function $f(x,k)$ will be henceforth denoted as scalar distribution and $\H_{rs}(x,k)$ 
as the spin density matrix in the phase space.

\section{Local equilibrium density operator}
\label{sec:ledo}

From the free Lagrangian for a neutral spin-1 particle \eq{Lag:free}, one can derive the canonical 
form of the energy-momentum and spin tensors
\begin{align}
\label{emt}
\opt{T}^{\m\n} &= - \opt{F}^{\m\r}\pt^\n \opt{A}_\r-\w^{\m\n}\left(-\frac{1}{4} \opt{F}_{\r\s}\opt{F}^{\r\s}+\frac{1}{2}m^2\opt{A}_\r \opt{A}^\r\right) \,,\\
\opt{S}^{\m\r\s} &= -\opt{F}^{\m\r}\opt{A}^{\s}+ \opt{F}^{\m\s}\opt{A}^{\r} \,.
\end{align}
The corresponding local equilibrium density operator (LEDO) describing a local Gibbs state specified by the thermal velocity $ \b_\m(x) $ and spin potential $ \O_{\r\s}(x) $ is given by~\cite{Zubarev:1979,Weert:1982,Becattini:2019dxo}
\begin{equation}
    \label{rho:LE}
    \opt{\r}_\LE = \frac{1}{Z_\LE}\exp\bigg\{-\int_{\S} d\S_\m(y) \ls \opt{T}^{\m\n}(y)\b_\n(y) -\frac{1}{2}\opt{S}^{\m\r\s}(y)\O_{\r\s}(y) \rs \bigg\} \,,
\end{equation}
where $ d\S_\m $ is the shorthand notation for $ d\S ~n_\m $ with $ d\S $ being the measure of the hypersurface $ \S$ and $ n_\m $ being its normal vector. $ Z_\LE $ is the local equilibrium partition function, which is the trace of the exponent term and ensures that the LEDO is properly normalized, $\Tr\opt{\r}=1$.
\begin{figure}[htbp]
    \centering
    \includegraphics[width=0.6\textwidth]{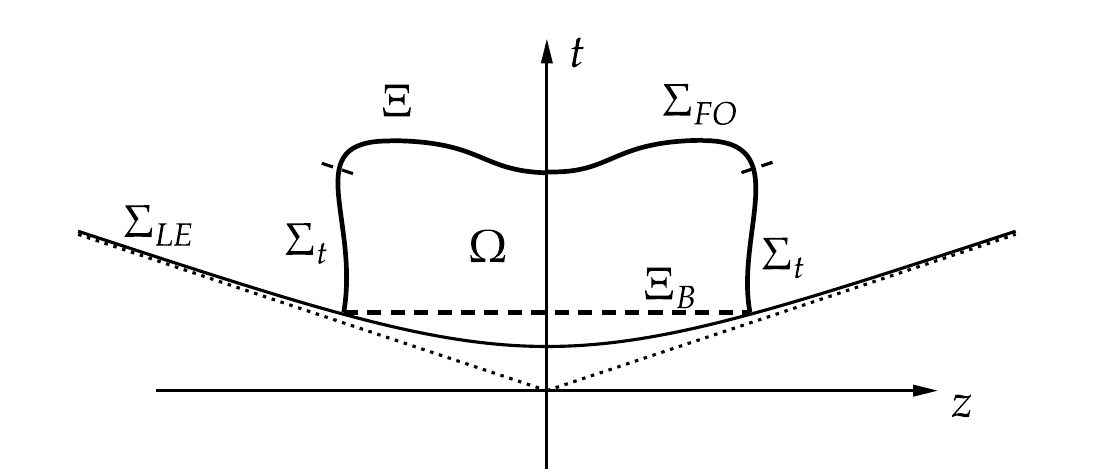}
    \caption{A sketch of the space-time diagram of a relativistic heavy ion collision. $ \S_{\LE} $ is the 3D space-like hypersurface where the local thermal equilibrium is achieved. $ \S_{\FO} $ is the freeze-out hypersurface, consisting of space-like hypersurface $ \Xi $ and time-like hypersurface $ \S_t $. In $ \O $, the volume surrounded by $ \S_{\LE} $ and $ \S_{\FO} $, the matter is in the QGP phase. $ \Xi_B $ is the hyperplane connecting the boundary of $ \S_{\FO} $.}
    \label{fig:surface1}
\end{figure}
Note that, in the LEDO, we treat the spin potential as an independent parameter. This is in line with the perspective of spin hydrodynamics~\cite{Florkowski:2017ruc,Hattori:2019lfp,Fukushima:2020ucl,Hongo:2021ona,Cao:2022aku,She:2021lhe,Gallegos:2021bzp} (for a review of spin hydrodynamics, see ref.~\cite{Huang:2024ffg}), where the spin polarization density relaxes to local equilibrium comparably slower
than momentum degrees of freedom~\cite{Hattori:2019lfp,Hongo:2021ona,Becattini:2018duy}. The 
spin potential will affect the spin alignment of vector bosons, as we will show.

The coupling field $ \b_\n(x),\,\O_{\r\s}(x) $ should be understood as Lagrange multipliers
for the maximization of entropy
\[
    S[\S] = -\Tr(\opt{\r}\ln{\opt{\r}}) \,,
\]
under the constraints
\begin{align}
    \label{con:t}
    n_\m T^{\m\n}(x) &= n_\m \Tr\lc\opt{T}^{\m\n}(x)\opt{\r}_\LE[\b,\O]\rc \,,\\
    \label{con:s}
    n_\m S^{\m\r\s}(x) &= n_\m \Tr\lc\opt{S}^{\m\r\s}(x)\opt{\r}_\LE[\b,\O]\rc \,,
\end{align}
where, on the left-hand side, there are the actual renormalized values of the currents.

The actual density operator in the Heisenberg picture must be static and $\opt{\r}_\LE$,
in general, is not. Nevertheless, the actual density operator of a system that achieves, at 
some time, local thermodynamic equilibrium can be expressed as the LEDO for some {\em fixed} hypersurface ($\Sigma_{LE}$ in \fig{fig:surface1}). 
By means of the Gauss theorem, the density operator can be rewritten as an exponential
of the 3D integral over the hypersurface at some ``current" time plus a 4D integral over
the encompassed region between the two 3D integrations \cite{Becattini:2019dxo}.
Provided that the thermodynamic fields $\beta$ and $\Omega$ are slowly varying, the
4D integral, which represents the dissipative contribution to any observable, can be 
neglected and one is left with the local equilibrium density operator at the current time; 
for relativistic heavy ion collisions, this is the freeze-out hypersurface $\Sigma_{FO}$.
Therefore, in practice, if the dissipative part is neglected, the calculation of mean
values can be carried out at the local thermodynamic equilibrium at the freeze-out by
replacing the hypersurface $\Sigma$ with $\Sigma_{FO}$ in \eq{rho:LE}.

In view of the above discussion, for a local Heisenberg operator $\opt{O}(x)$, such as 
$\opt{f}_{rs}(x,k)$,  we have its ensemble average given by
\begin{equation}
    \label{ensembleave}
    O(x) \equiv \lan\opt{O}(x)\ran \equiv \Tr[\opt{\r}_{\rm LE}\opt{O}(x)] = \frac{1}{Z_{\rm LE}}\Tr\ls e^{\opt{A}+\opt{B}}\opt{O}(x)\rs,
\end{equation}
where we have introduced the abbreviations
\begin{align}
    \opt{A} &= -\opt{P}^\m\b_\m(x) \,,\\
    \opt{B} &= -\int_{\S_{\FO}} d\S_\m(y) \lb \opt{T}^{\m\n}(y)\D\b_\n(y)-\frac{1}{2}\opt{S}^{\m\r\s}(y)\O_{\r\s}(y)\rb
\end{align}
with $\opt{P}^\m=\int_{\S_{\FO}} d\S_\n(y)\opt{T}^{\n\m}(y)$ and $\D\b_\m(y)=\b_\m(y)-\b_\m(x)$.

In the following, we assume that $\D\b_\m(y)$ is small on $\S_\text{FO}$, which is justified when $\b_\m$ varies slowly within the correlation lengths between $\opt{O}$ and the operator $\opt{T}^{\m\n}$. Meanwhile, we assume that $\O_{\r\s}(y)$ is also small on the hypersurface $\S_\text{FO}$, which could be of the same order as $\D\b_\m(y)$ (this is supported by the fact that the observed spin polarization of hadrons is small). Thus \eq{ensembleave} can be calculated in a perturbative way in derivative expansion using the power counting rules
\begin{align}
    \D\b_\n(y) &= (y-x)^{\a_1}\pt_{\a_1}\b_\n(x) +\frac{1}{2}(y-x)^{\a_1}(y-x)^{\a_2}\pt_{\a_1}\pt_{\a_2}\b_\n(x)+\cdots = \mathcal{O}(\pt) \,,\\
    \O_{\r\s}(y) &= \O_{\r\s}(x) +(y-x)^{\a_1}\pt_{\a_1}\O_{\r\s}(x) + \cdots = \mathcal{O}(\pt) \,.
\end{align}
It should be emphasized that in the current content, the gradient expansion is equivalent to the $\hbar$ expansion.

Using the cumulant expansion
\begin{align}
    e^{\opt{A}+\opt{B}} &= e^{\opt{A}}\sum_{n=0}^\infty \opt{B}_n \,, \qquad \opt{B}_0 =1 \,,\\
    \opt{B}_n &= \int^1_0 d\l_1 \int^{\l_1}_0 d\l_2 \cdots \int^{\l_{n-1}}_0 d\l_n\opt{B}(\l_1)\cdots\opt{B}(\l_n),\quad n\geq 1
\end{align}
with $ \opt{B}(\l_i)\equiv e^{-\l_i \opt{A}} \opt{B}e^{\l_i \opt{A}} $, we rewrite \eq{ensembleave} as
\begin{equation}
    \label{exp:expansion}
    O(x) = \frac{\sum_{n=0}^\infty \lan\opt{B}_n \opt{O}(x)\ran_0}{\sum_{n=0}^\infty \lan\opt{B}_n \ran_0} \,,
\end{equation}
where $ \lan\cdots\ran_0 $ denotes the expectation value with density operator $ \opt{\r}_0 = e^{\opt{A}}/Z_0$ with $Z_0=\Tr e^{\opt{A}} $. In our power counting scheme, $\opt{A}$ is $\mathcal{O}(1)$ and $\opt{B}$ contains an $\mathcal{O}(\pt)$ term as well as higher order terms  $\mathcal{O}(\pt^n)$ with $n=2,3,4,...$, therefore we find that the lowest order part of $\opt{B}_n$ is $\mathcal{O}(\pt^n)$. This allows us to expand the ensemble average $ O(x) $ in powers of $ \pt $
\begin{equation}
    \label{ensembleave3}
    O(x) = O^{(0)}(x) +O^{(1)}(x) +O^{(2)}(x) +\mathcal{O}(\pt^3) \,,
\end{equation}
where we denote
\begin{align}
    \label{ensembleave4}
    O^{(0)}(x) &= \lan\opt{O}(x)\ran_{0} \equiv\frac{1}{Z_0}\Tr\lb e^{\opt{A}}\opt{O}(x)\rb \,,\\
    O^{(1)}(x) &= \lan\opt{B}_1\opt{O}(x)\ran_{0,c} \equiv\lan\opt{B}_1\opt{O}(x)\ran_0 -\lan \opt{B}_1\ran_0 O^{(0)}(x) \,,\\
        \label{ensembleave444}
        O^{(2)}(x) &= \lan\opt{B}_2\opt{O}(x)\ran_{0,c} \equiv\lan\opt{B}_2\opt{O}(x)\ran_0 -\lan \opt{B}_2\ran_0 O^{(0)}(x) -\lan \opt{B}_1\ran_0 O^{(1)}(x) \,.
\end{align}
In the above expressions, $\lan\cdots\ran_{0,c}$ represents the connected part of the correlator.
The term ``connected" is not entirely coincidental. One may notice that \eq{exp:expansion} resembles Feynman's path integral formula in perturbative QFT, where terms in the denominator are called vacuum fluctuations and can cancel out the vacuum fluctuations in the numerator.

We note that $ \opt{B}_n $  contains not only $ \mathcal{O} (\pt^n) $ terms but also $ \mathcal{O} (\pt^{n^\prime}) $ terms with $n^\prime=n+1,n+2,\cdots$. For clarification, we introduce operator $ \opt{\mathcal{B}}_n $ to represent precisely the $ \mathcal{O} (\pt^n) $ terms in $ \opt{B}_n $. The relations \eqs{ensembleave3}{ensembleave444} remain valid for $ \opt{\mathcal{B}}_n $. The leading two $ \opt{\mathcal{B}}_n $'s are given by
\small{
\begin{align}
    \label{def:B1}
    \opt{\mathcal{B}}_1 \equiv& \int_0^1 d\l_1 \int_{\S_{\FO}} d\S_{\m_1}(y_1) \ls -\pt_{\a_1}\b_{\n_1}(x)(y_1-x)^{\a_1}\opt{T}^{\m_1\n_1}(y_1^{(\b)}) +\frac{1}{2}\O_{\r_1\s_1}(x)\opt{S}^{\m_1\r_1\s_1}(y_1^{(\b)})\rs,\\
    \label{def:B2}
    \begin{split}
    \opt{\mathcal{B}}_2 \equiv& \int_0^1 d\l_1 \int_0^{\l_1} d\l_2 \int_{\S_{\FO}} d\S_{\m_1}(y_1) d\S_{\m_2}(y_2)\\
    & \times\left[\pt_{\a_1}\b_{\n_1}(x)\pt_{\a_2}\b_{\n_2}(x)(y_1-x)^{\a_1}(y_2-x)^{\a_2}\opt{T}^{\m_1\n_1}(y_1^{(\b)})\opt{T}^{\m_2\n_2}(y_2^{(\b)})\right.\\
    & -\frac{1}{2}\pt_{\a_1}\b_{\n_1}(x)\O_{\r_2\s_2}(x)(y_1-x)^{\a_1}\opt{T}^{\m_1\n_1}(y_1^{(\b)})\opt{S}^{\m_2\r_2\s_2}(y_2^{(\b)})\\
    & -\frac{1}{2}\O_{\r_1\s_1}(x)\pt_{\a_2}\b_{\n_2}(x)(y_2-x)^{\a_2}\opt{S}^{\m_1\r_1\s_1}(y_1^{(\b)})\opt{T}^{\m_2\n_2}(y_2^{(\b)})\\
    & \left.+\frac{1}{4}\O_{\r_1\s_1}(x)\O_{\r_2\s_2}(x)\opt{S}^{\m_1\r_1\s_1}(y_1^{(\b)})\opt{S}^{\m_2\r_2\s_2}(y_2^{(\b)})\right]\\
    & +\int_0^1 d\l_1 \int_{\S_\FO} d\S_{\m_1}(y_1)\left[-\pt_{\a_1}\pt_{\a_2}\b_{\n_1}(x)\frac{1}{2}(y_1-x)^{\a_1}(y-x)^{\a_2}\opt{T}^{\m_1\n_1}(y_1^{(\b)})\right.\\
    & +\left.\frac{1}{2}\pt_{\a_1}\O_{\r_1\s_1}(x)(y_1-x)^{\a_1}\opt{S}^{\m_1\r_1\s_1}(y_1^{(\b)})\right]
    \end{split}
\end{align}
}
with $ y_i^{(\b)} = y_i-i\l_i\b(x) $. Such a shift in position is obtained from
\[
    e^{\l_i \b(x)\cdot\opt{P}}\opt{B}(y_i)e^{-\l_i\b(x)\cdot\opt{P}} = \opt{B}(y_i-i\l_i\b(x)) \,,
\]
which is because $ \opt{P}_\m $ generates space-time translation for operators in the Hilbert space.

\section{From MVSD to polarization}
\label{sec:MVSDtop}

In a system described by the LEDO in \eq{rho:LE}, the actual spin density matrix as a function 
of the momentum in \eq{eq:cooper-frye} is most conveniently expressed as an integral over 
the freeze-out hypersurface (we remind that $\Sigma$ is arbitrary) because this is the most convenient 
choice to expand the thermodynamic fields. By using the decomposition \eq{fdecomp}
we thus have
\begin{equation}
	\label{eq:cooperfrye}
    \H_{rs}(k) = \frac{\int_{\S_\FO} d\S\cdot k\,f_{rs}(x,k)}{\int_{\S_\FO} d\S\cdot k\,f(x,k)}
    = \frac{\int_{\S_\FO} d\S\cdot k\,\Theta_{rs}(x,k) f(x,k)}{\int_{\S_\FO} d\S\cdot k\,f(x,k)} \,.
\end{equation}
By using the definitions \eq{exp:P} and \eq{exp:T1}, we then have
\begin{align}\label{eq:cooperfrye2}
    S_\text{rest}^{i}(k) &= \frac{\int_{\S_\FO} d\S\cdot k\,S_\text{rest}^{i}(x,k)f(x,k)}{\int_{\S_\FO} d\S\cdot k\,f(x,k)} \,,\\
\label{eq:cooperfrye3}
    \fT_{2,m}(k) &= \frac{\int_{\S_\FO} d\S\cdot k\,\fT_{2,m}(x,k)f(x,k)}{\int_{\S_\FO} d\S\cdot k\,f(x,k)} \,,
\end{align}
where the integrands are obtained by tracing the spin density matrix in phase space instead of the
properly defined one, in \eq{exp:P} and \eq{exp:T1}.

Replacing the spin density matrix with the spin density matrix in phase space in \eq{exp:P} and \eq{exp:T2}
we obtain
\begin{align}
\label{polarization-Px}
        S_\text{rest}^1(x,k) &= \frac{1}{f(x,k)} \sum_{i=0}^{\infty} 2\sqrt{2}\Im f^{(2i+1)}_{01}(x,k) \,,\\
        S_\text{rest}^2(x,k) &= \frac{1}{f(x,k)} \sum_{i=0}^{\infty}2f^{(2i+1)}_{11}(x,k) \,,\\
        S_\text{rest}^3(x,k) &= \frac{1}{f(x,k)} \sum_{i=0}^{\infty} 2\sqrt{2}\Re f^{(2i+1)}_{01}(x,k) \,,\\
        \fT_{2,0}(x,k) &= \frac{1}{f(x,k)} \sum_{i=0}^{\infty}\sqrt{\frac{2}{3}}\ls f^{(2i+2)}_{11}(x,k)-f^{(2i+2)}_{00}(x,k)\rs \,,\\
        \fT_{2,1}(x,k) = -\fT^{*}_{2,-1}(x,k) &= \frac{1}{f(x,k)} \sum_{i=0}^{\infty} (-\sqrt{2})f^{(2i+2)}_{01}(x,k) \,,\\
\label{polarization-T22}
        \fT_{2,2}(x,k) = \fT^{*}_{2,-2}(x,k) &= \frac{1}{f(x,k)} \sum_{i=0}^{\infty} f^{(2i+2)}_{-11}(x,k) \,,
\end{align}
where $ f(x,k) = \sum_{i=0}^{\infty} [2f_{11}^{(2i)}(x,k) +f_{00}^{(2i)}(x,k)]$ because $f_{-1-1}^{(2i)}(x,k)=f_{11}^{(2i)}(x,k)$ and $ \, $ $f_{rr}^{(2i+1)}(x,k)=-f_{-r-r}^{(2i+1)}(x,k)$ (see \eq{eq:PTrev}). Here $f_{rs}^{(i)}$ denotes the $\mathcal{O}(\pt^i)$  part of the whole MVSD. 
The MVSD at zeroth order in gradients should be $f_{rs}^{(0)}= f_{00}^{(0)}\delta_{rs} $ . Note that, in \eqs{polarization-Px}{polarization-T22}, the even (odd) order terms do not contribute to the numerators of vector (tensor) polarization. This is due to the following space-time reversal property of MVSD (For details of the derivation, see appendix~\ref{sec:PTrev})
\begin{equation}
\label{eq:PTrev}
    f^{(n)}_{r,s}(x,k) = (-1)^{r+s+n} f^{(n)}_{-s,-r}(x,k) \,.
\end{equation}
The leading order contributions to the vector and tensor polarization are
\begin{align}
\label{leading:P}
    \lc S_\text{rest}^1,S_\text{rest}^2,S_\text{rest}^3\rc &\approx \frac{1}{f^{(0)}}
    \lc 2\sqrt{2}\Im f^{(1)}_{01},\,2f^{(1)}_{11},\,2\sqrt{2}\Re f^{(1)}_{01}\rc \,,\\
    \lc \fT_{2,0},\fT_{2,1},\fT_{2,2}\rc &\approx \frac{1}{f^{(0)}}\lc\sqrt{\frac{2}{3}}(f^{(2)}_{11}-f^{(2)}_{00}),\,-\sqrt{2}f^{(2)}_{01},\,f^{(2)}_{-11}\rc \,,
\label{leading:T}
\end{align}
where $ f^{(0)} $ is the zeroth order scalar distribution $ f^{(0)} = \sum_{r=0,\pm1} f_{rr}^{(0)} = 3 f_{00}^{(0)}$.
Accordingly, the spin alignment in phase space is given by
\begin{equation}
\label{eq:spin-align}
    \H_{00}(x,k)-\frac{1}{3} \approx \frac{2(f^{(2)}_{00}-f^{(2)}_{11})}{3f^{(0)}}.
\end{equation}
Therefore, the evaluation of \eq{eq:spin-align} does require the expansion of the MSVD at the second order.

\section{Zeroth- and first-order results}
\label{sec:zeroth:dis}

In this section, we focus on the zeroth and first-order parts of the MVSD. The zeroth order part can be easily calculated using $ \lan \opt{a}^{r\dg}_\bk\opt{a}^{s}_\bq\ran_0 = (2E_\bk)(2\p)^3\d^{(3)}(\bk-\bq) \d^{rs} n_B(\b(x)\cdot k)$. The result reads
\begin{equation}
    \label{frszero}
    {f}_{rs}^{(0)}(x,k) = \d(k^2-m^2)\h(k^0) \d_{rs} n_B(\b(x)\cdot k) \,,
\end{equation}
where $ n_B(\b(x)\cdot k) $ is the Bose-Einstein distribution, \ie, $ 1/(e^{\b(x)\cdot k}-1) $.

For the MVSD at first order in gradients, we split it into two parts as $f^{(1)}_{rs}=f^{(1)}_{rs}|_{T}+f^{(1)}_{rs}|_{S}$ with
\begin{align}
\label{OTJS1}
    f^{(1)}_{rs}|_{T} &\equiv -\int_0^1 d\l_1 \int_{\S_{\FO}} d\S_{\m_1}(y_1) [\pt_{\a_1}\b_{\n_1}(x)] (y_1-x)^{\a_1} \lan\opt{T}^{\m_1\n_1}(y_1^{(\b)})\opt{f}_{rs}(x,k)\ran_{0,c}\,,\quad\\
\label{OTJS2}
    f^{(1)}_{rs}|_{S} &\equiv \int_0^1 d\l_1 \int_{\S_{\FO}} d\S_{\m_1}(y_1) \O_{\r_1\s_1}(x) \lan\frac{1}{2}\opt{S}^{\m_1\r_1\s_1}(y_1^{(\b)})\opt{f}_{rs}(x,k)\ran_{0,c}\,,
\end{align}
where $f^{(1)}_{rs}|_{T}$ ($f^{(1)}_{rs}|_{S}$) represents the contribution from the correlation between the energy-momentum operator
(the spin operator) and the MVSD operator. By substituting eqs.~\eqref{def:MVSD}, \eqref{def:Wig}, \eqref{emt}, and the quantized field operator \eq{proca:field:decomp} into  $f^{(1)}_{rs}|_{T}$, we obtain 
{\small
\begin{equation}
\label{frsT1}
    \begin{split}
    f^{(1)}_{rs}|_{T} =& \e^{\g_3*}_{r}(k)\e^{\g_0}_{s}(k)\int_0^1 d\l_1 \int_{\S_{\FO}} d\S_{\m_1}(y_1) \ls\pt_{\a_1}\b_{\n_1}(x)\rs (y_1-x)^{\a_1}\\
    & \times \prod_{i=0}^{3}\lb\sum_{a_i=-1}^{1}\int\idp{3}{\bp_i}\frac{1}{2E_{\bp_i}}\rb (2\p)^3 \d^{(4)}\lb k-\frac{p_1}{2}-\frac{p_2}{2}\rb \Big\{\Big[p_1^{\m_1} p_2^{\n_1}\w^{\g_1\g_2} \\
    & -p_1^{\g_2}p_2^{\n_1}\w^{\m_1\r_1} -\frac{1}{2}\lb p_1\cdot p_2\rb\w^{\m_1\n_1}\w^{\g_1\g_2} +\frac{1}{2}\w^{\m_1\n_1}\left(p_1^{\g_2} p_2^{\g_1} +m^2\w^{\g_1\g_2}\right)\Big] e^{i(p_0-p_3)\cdot x}\\
    & \times\Big[\lan \opt{a}_{\bp_1}^{a_1} \opt{a}^{a_0\dg}_{\bp_0}\ran_0 \lan\opt{a}_{\bp_2}^{a_2\dg}\opt{a}^{a_3}_{\bp_3}\ran_0 \e_{\g_0}^{a_0*}(p_0)\e_{\g_1}^{a_1}(p_1)\e_{\g_2}^{a_2*}(p_2)\e_{\g_3}^{a_3}(p_3) e^{-i(p_1-p_2)\cdot(y_1-i\l_1\b(x))}\\
    & +\lan \opt{a}_{\bp_1}^{a_1\dg} \opt{a}^{a_3}_{\bp_3} \ran_0 \lan \opt{a}_{\bp_2}^{a_2} \opt{a}^{a_0\dg}_{\bp_0} \ran_0 \e_{\g_0}^{a_0*}(p_0)\e_{\g_1}^{a_1*}(p_1)\e_{\g_2}^{a_2}(p_2)\e_{\g_3}^{a_3}(p_3) e^{i(p_1-p_2)\cdot(y_1-i\l_1\b(x))}\Big]\Big\}.
    \end{split}
\end{equation}
}

For the integration over $\S_{\FO} $ in the Cooper-Frye type formulas, we consider a scenario that $ \Xi $, the space-like part of $\S_{\FO}$ (see \fig{fig:surface1}),
is a flat hyperplane and is sufficiently large, while the contribution from $\S_t$, the time-like part of $\S_{\FO} $, can be neglected.
If these conditions are relaxed, $\Xi$'s non-planarity and the evaluation of $ f_{rs}(x,k) $ with $ x \in \S_t $
will bring additional complexity to our calculation (see appendix~\ref{sec:time-like}), which will be studied in a separate paper.
Based on the above assumptions, for the integration over $\S_{\FO} $ in an expression like \eq{frsT1}, we can replace $ \S_\FO $ by hyperplane $ \Xi $,
which can be verified in appendix~\ref{sec:time-like}.
Without loss of generality, in the following, we take the hyperplane $ \Xi $ to be perpendicular to the time direction
\footnote{Due to the Lorentz covariance of the Wigner function, the coordinate time direction, $ \htt^{\m} $, in our expressions
can always be replaced by the normal vector for a general flat hyperplane.}, $ \htt^{\m} = (1,\bm 0) $, namely the isochronous freeze-out
\footnote{If the integration over the four-volume $\O$ in \fig{fig:surface1} does not contribute to spin polarization~\cite{Becattini:2021suc},
we can replace the integration over $\S_\FO$ by the integration over $\Xi_B$.
Our results can be applied to this scenario as well by simply replacing $\Xi$ by $\Xi_B$.}.
This leads to
\begin{equation}
\label{hypersurface-int}
\begin{split}
    & \int_{\Xi} d\Xi_{\m_1}(y_1)\prod_{i=1}^{2}\lb\int\idp{3}{\bp_i}\rb(y_1-x)^{\a_1} e^{-i(p_1-p_2)\cdot (y_1-x)} \\
    =& \htt_{\m_1} \prod_{i=1}^{2}\lb\int\idp{3}{\bp_i}\rb (2\p)^3 \d^{(3)}\lb\bp_1-\bp_2\rb \lb-\frac{i}{2}\rb \lb\na_{1}^{\a_1} -\na_{2}^{\a_1}\rb,
\end{split}
\end{equation}
where the transverse derivative operator is defined as $\na_i^{\a_j}\equiv\pt_i^{\a_j}-\hat{t}^{\a_j}(\hat{t}\cdot\pt_i)$  with  $\pt_{i}^{\a_j}= \pt/\pt p_{i,\a_j} $ being the covariant momentum derivative vector. Noting that $\na_i^0$ is always zero because the four-momentum $p_i^\mu$ is restricted on the normal mass-shell with $p_i^0=\sqrt{\bp_i^2+m^2}$, indicating that $p_i^0$ is not an independent parameter. Therefore, for $ \a_1=0 $, both sides of the equation are identically zero.
When calculating \eq{frsT1}, we first do the integration over $ \bp_0 $ and $ \bp_3 $, then an integral by part shown in \eq{hypersurface-int}. We note that the momentum derivatives in \eq{hypersurface-int} also act on $ \d^{(4)}\lb k-\frac{p_1}{2}-\frac{p_2}{2}\rb $. However, one can check that the delta function commutes with the partial derivatives, which allows us to move it outside of the momentum derivatives.
Using $ \lan\opt{a}^{r\dg}_\bk\opt{a}^{s}_\bq\ran_0 = (2E_\bk) (2\p)^3 \d^{(3)}(\bk-\bq) \d^{rs} n_B $ and $ \lan\opt{a}^{r}_\bk\opt{a}^{s\dg}_\bq\ran_0 = (2E_\bk) (2\p)^3\d^{(3)}(\bk-\bq) \d^{rs} (1+n_B) $, \eq{frsT1} is simplified to the following form
\begin{equation}
    \label{frsT2}
    f^{(1)}_{rs}|_{T}(x,k) = \frac{i}{2} \d(k^2-m^2)\h(k^0) n_B(1+n_B) \e^{\g_3*}_{r}(k)\e^{\g_0}_{s}(k) \lc\htt_{[\g_0}[\pt_{\g_3]}\b_{\n_1}(x)]\hat{k}^{\n_1}\rc
\end{equation}
with $ n_B $ standing for $ n_B(\b(x)\cdot k) $.
(The detailed derivation of \eq{frsT2} is tedious but straightforward. We first do the integration over $ \bp_0,\bp_1,\bp_2,\bp_3 $, then the integration over $\l_1$.)
The calculation of $ f_{rs}^{(1)}|_S $ is completely the same, which gives
\begin{equation}
\label{frsS2}
    f^{(1)}_{rs}|_{S}(x,k) = \frac{i}{2} \d(k^2-m^2)\h(k^0) n_B(1+n_B) \e^{\g_3*}_{r}(k)\e^{\g_0}_{s}(k) \lc 2\O_{\g_0\g_3} +\hat{k}^{\r_1}\O_{\r_1[\g_0}\htt_{\g3]}\rc \,.
\end{equation}
As the terms in the curly braces of \eqs{frsT2}{frsS2} are both antisymmetric w.r.t. $ \g_0 \leftrightarrow \g_3 $, we have
\begin{equation*}
    f_{rs}^{(1)}(x,k) = -f_{-s,-r}^{(1)}(x,k) \,,
\end{equation*}
which agrees with \eq{eq:PTrev}.
Such a relation implies that $f_{00}^{(1)}(x,k)=0$ and $f_{11}^{(1)}(x,k)+f_{-1,-1}^{(1)}(x,k)=0$, leading to a vanishing spin alignment at the first order in gradients, which agrees with the results in refs.~\cite{Yang:2017sdk,Sheng:2019kmk}. 

\section{Diagrammatic scheme}
\label{sec:diagram}
In order to derive a formula for the spin alignment, we need to evaluate the second-order MVSD since the zeroth- and first-order results do not yield a tensor polarization (see section~\ref{sec:MVSDtop}).
For this purpose, we develop a diagrammatic scheme analogous to the Feynman diagram to calculate $ f_{rs}(x,k) $ order by order in space-time gradients.

Similar to \eqs{OTJS1}{OTJS2}, we split the second-order MVSD into six terms, respectively related to the six terms of $ \opt{\mathcal{B}}_2 $ given in \eq{def:B2}, 
\begin{equation}
    \label{expand:f2}
    f^{(2)}_{rs} \equiv \lan \opt{\mathcal{B}}_2 \opt{f}_{rs} \ran_{0,c} = f^{(2)}_{rs}|_{TT} +f^{(2)}_{rs}|_{TS} +f^{(2)}_{rs}|_{ST} +f^{(2)}_{rs}|_{SS} +f^{(2)}_{rs}|_{T} +f^{(2)}_{rs}|_{S} \,,
\end{equation}
where ``$T$'' and ``$S$'' represent the energy-momentum and spin operators, respectively.
The term $ f^{(2)}_{rs}|_{TT} $, for example, denotes the contribution from the correlation between two energy-momentum operators and the MVSD operator.
Another term, $f^{(2)}_{rs}|_{TS}$, is the contribution from the correlation between one energy-momentum operator, one spin operator, and the MVSD operator.
The other terms can be understood similarly. Therefore, we would expect (and we will show) that $ f^{(2)}_{rs}|_{TT}\propto (\pt\b)(\pt\b)$,
$ f^{(2)}_{rs}|_{TS}\propto (\pt\b)\O$, $ f^{(2)}_{rs}|_{ST}\propto \O(\pt\b)$, $ f^{(2)}_{rs}|_{SS}\propto \O \O$,
$ f^{(2)}_{rs}|_{T}\propto \pt\pt\b$, and $ f^{(2)}_{rs}|_{T}\propto \pt\O$.

Let us take $ f^{(2)}_{rs}|_{TT} $ as an example. It contains a term as follows
\begin{equation}
\label{example:TT}
    \begin{split}
    & \,\e^{\g_5*}_{r}(k)\e^{\g_0}_{s}(k)\ls\pt_{\a_1}\b_{\n_1}(x)\rs\ls\pt_{\a_2}\b_{\n_2}(x)\rs \int_0^1d\l_1 \int_0^{\l_1}d\l_2 \prod_{i=0}^{5}\lb\sum_{a_i=-1}^{+1}\int\idp{3}{\bp_i}\rb\\
    & \times \int_{\Xi} d\Xi_{\m_1}(y_1)d\Xi_{\m_2}(y_2) (y_1-x)^{\a_1}(y_2-x)^{\a_2} e^{i\,(p_0-p_5)\cdot x}e^{-i(p_1+p_2)\cdot y_1} \\
    & \times e^{+i(p_3+p_4)\cdot y_2} (2\p)^3 \d^{(4)}\lb k-\frac{p_0}{2}-\frac{p_5}{2}\rb t\,^{\m_1\n_1}_{(1)~\g_1\g_2}(p_1,p_2) t\,^{\m_2\n_2}_{(2)~\g_3\g_4}(-p_3,-p_4) \\
    & \times \e_{\g_0}^{a_0*}(p_0)\e^{\g_1}_{a_1}(p_1)\e^{\g_2}_{a_2}(p_2)\e^{\g_3*}_{a_3}(p_3)\e^{\g_4*}_{a_4}(p_4)\e_{\g_5}^{a_5}(p_5)\lan\opt{a}^{a_1}_{\bp_1}\opt{a}^{a_2}_{\bp_2}\opt{a}^{a_3\dg}_{\bp_3}\opt{a}^{a_4\dg}_{\bp_4}\opt{a}^{a_0\dg}_{\bp_0}\opt{a}^{a_5}_{\bp_5}\ran_{0,c} \,,
    \end{split}
\end{equation}
where we denote
\begin{equation}
    \begin{split}
    t\,^{\m_1\n_1}_{(1)~\g_1\g_2}(p_1,p_2) =& \frac{e^{-\l_1(p_1+p_2)\cdot \b}}{2E_{\bp_1}2E_{\bp_2}} \Big[p_1^{\m_1}p_2^{\n_1}\w_{\g_1\g_2} -{p_1}_{\g_2} p_2^{\n_1}\w^{\m_1}_{~~\g_1}\\
    & -\frac{1}{2}(p_1\cdot p_2)\w_{\g_1\g_2}\w^{\m_1\n_1} +\frac{1}{2}{p_1}_{\g_2} {p_2}_{\g_1}\w^{\m_1\n_1} -\frac{1}{2}m^2\w_{\g_1\g_2}\w^{\m_1\n_1}\Big] \,,
    \end{split}
\end{equation}
which comes from $ \opt{T}^{\m_1\n_1}(x) $; $t\,^{\m_2\n_2}_{(2)~\g_3\g_4}(p_3,p_4)$ is completely the same but from $ \opt{T}^{\m_2\n_2}(x) $.
We emphasize that the expression \eqref{example:TT} is just one part of the whole  $ f^{(2)}_{rs}|_{TT} $.
The full result for $ f^{(2)}_{rs}|_{TT} $ will be shown in section~\ref{sec:2nd:order} with a diagrammatic scheme.

Utilizing the Wick theorem, we obtain
\begin{equation}
\label{thm:wick}
    \begin{split}
    &\lan \opt{a}^{a_1}_{\bp_1}\opt{a}^{a_2}_{\bp_2}\opt{a}^{a_3\dg}_{\bp_3}\opt{a}^{a_4\dg}_{\bp_4}\opt{a}^{a_0\dg}_{\bp_0}\opt{a}^{a_5}_{\bp_5}\ran_{0,c}\\
    =& \lan\opt{a}^{a_1}_{\bp_1}\opt{a}^{a_0\dg}_{\bp_0}\ran_0 \lan\opt{a}^{a_2}_{\bp_2}\opt{a}^{a_3\dg}_{\bp_3}\ran_0 \lan\opt{a}^{a_4\dg}_{\bp_4}\opt{a}^{a_5}_{\bp_5}\ran_0 +\lan\opt{a}^{a_1}_{\bp_1}\opt{a}^{a_4\dg}_{\bp_4}\ran_0 \lan\opt{a}^{a_2}_{\bp_2}\opt{a}^{a_0\dg}_{\bp_0}\ran_0 \lan\opt{a}^{a_3\dg}_{\bp_3}\opt{a}^{a_5}_{\bp_5}\ran_0\\
    &+ \lan\opt{a}^{a_1}_{\bp_1}\opt{a}^{a_3\dg}_{\bp_3}\ran_0 \lan\opt{a}^{a_2}_{\bp_2}\opt{a}^{a_0\dg}_{\bp_0}\ran_0 \lan\opt{a}^{a_4\dg}_{\bp_4}\opt{a}^{a_5}_{\bp_5}\ran_0 +\lan\opt{a}^{a_1}_{\bp_1}\opt{a}^{a_0\dg}_{\bp_0}\ran_0 \lan\opt{a}^{a_2}_{\bp_2}\opt{a}^{a_4\dg}_{\bp_4}\ran_0 \lan\opt{a}^{a_3\dg}_{\bp_3}\opt{a}^{a_5}_{\bp_5}\ran_0 \,.
    \end{split}
\end{equation}
We can use diagrams to express these different terms. For instance, the first term is represented by
\begin{equation}
    \lan\opt{a}^{a_1}_{\bp_1}\opt{a}^{a_0\dg}_{\bp_0}\ran_0 \lan\opt{a}^{a_2}_{\bp_2}\opt{a}^{a_3\dg}_{\bp_3}\ran_0 \lan\opt{a}^{a_4\dg}_{\bp_4}\opt{a}^{a_5}_{\bp_5}\ran_0 =
\begin{tikzpicture}[baseline=-0.65ex]
    \node[TSnode]   (T1)    [label=70:$ p_1 $,
                            label=-70:$ p_2 $,
                            label=110:$ a_1 $,
                            label=-110:$ a_2 $]                     {};
    \node[TSnode]   (T2)    [right=of T1,
                            label=70:$ p_3 $,
                            label=-70:$ p_4 $,
                            label=110:$ a_3 $,
                            label=-110:$ a_4 $]                     {};
    \node[fnode]    (f)     [right=of T2,
                            label=70:\small{$ p_0 $},
                            label=-70:\small{$ p_5 $},
                            label=110:$ a_0 $, label=-110:$ a_5 $]      {};
    \draw[->-]  (f.north)     to [out=90, in=90]      (T1.north);
    \draw[-<-]  (T1.south)    to [out=-90, in=90]     (T2.north);
    \draw[->-]  (T2.south)    to [out=-90, in=-90]    (f.south);
\end{tikzpicture},
\end{equation}
where the left arrow ``$ a_1,\,p_1\leftarrow a_2,\,p_2 $" represents $ \lan\opt{a}^{a_1}_{\bp_1}\opt{a}^{a_2\dg}_{\bp_2}\ran_0 $
and the right arrow ``$ a_1,\,\bp_1\rightarrow a_2,\,\bp_2 $" represents $ \lan \opt{a}^{a_1\dg}_{p_1}\opt{a}^{a_2}_{p_2}\ran_0 $\,.
Since a mean value like $ \lan\opt{a}^{a_1}_{\bp_1}\opt{a}^{a_2\dg}_{\bp_2}\ran_0 $ is proportional to $\delta_{a_1a_2}\delta^{(3)}(\bp_1-\bp_2)$,
the momenta and the spin indices connected by a line should be equal, respectively.
On the other hand, the diagram element
\begin{equation}
\begin{tikzpicture}[baseline=-0.65ex]
    \node[TSnode]   (T1)    [label=70:$ p_1 $,
                            label=-70:$ p_2 $,
                            label=110:$ a_1 $,
                            label=-110:$ a_2 $]                     {};
    \node           (up)    [above=of T1]{};
    \node           (down)  [below=of T1]{};
    \draw[-<-]      (T1)    to [out=90,in=-90]          (up);
    \draw[-<-]      (T1)    to [out=-90,in=90]          (down);
\end{tikzpicture}
\end{equation}
represents any c-number function that multiplying with $ \opt{a}^{a_1}_{\bp_1}\opt{a}^{a_2}_{\bp_2} $.
The arrow going into the box represents the annihilator $ \opt{a} $, while the arrow going out of the box represents creator $ \opt{a}^{\dg} $.

Furthermore, we can generalize the rules of the diagram to incorporate other terms in the expression \eqref{example:TT}.
We introduce some details to the diagram, namely
\begin{equation}
\begin{tikzpicture}[baseline=-0.65ex]
    \node[TSnode]   (T1)    [label={[label distance=0.1cm] 70:$p_1$},
                            label={[label distance=0.1cm]-70:$p_2$},
                            label={[label distance=0.1cm] 110:$a_1$},
                            label={[label distance=0.1cm]-110:$a_2$},
                            label={[label distance=0.3cm] 90:$\g_1$},
                            label={[label distance=0.3cm]-90:$\g_2$}]                   {$D_{(1)}^{\a_1}T_{(1)}^{\m_1\n_1}$};
    \node[TSnode]   (T2)    [right=of T1,
                            label={[label distance=0.1cm] 70:$p_3$},
                            label={[label distance=0.1cm]-70:$p_4$},
                            label={[label distance=0.1cm] 110:$a_3$},
                            label={[label distance=0.1cm]-110:$a_4$},
                            label={[label distance=0.3cm] 90:$\g_3$},
                            label={[label distance=0.3cm]-90:$\g_4$}]                   {$D_{(2)}^{\a_2}T_{(2)}^{\m_2\n_2}$};
    \node[fnode]    (f)     [right=of T2,
                            label=70:\small{$ p_0 $},
                            label=-70:\small{$ p_5 $},
                            label={[label distance=0.3cm] 90:$\g_0$},
                            label={[label distance=0.3cm]-90:$\g_5$},
                            label=110:$ a_0 $, label=-110:$ a_5 $]      {$f$};

    \draw[->-]  (f.north)     to [out=90, in=90]      (T1.north);
    \draw[-<-]  (T1.south)    to [out=-90, in=90]     (T2.north);
    \draw[->-]  (T2.south)    to [out=-90, in=-90]    (f.south);
\end{tikzpicture}.
\end{equation}

We clarify the meaning of various symbols in the following. By comparing with \eq{example:TT}, we can infer that each line entering (exiting) the boxes should carry the polarization vector (the conjugate of polarization vector), namely
\begin{equation}
\begin{tikzpicture}[baseline=-0.65ex]
    \node[fTSnode]   (T1)    [label= 70:$p_i$,
                            label= 110:$a_i$,
                            label={[label distance=0.4cm] 93:$\g_i$}]          {};
    \node           (up1)   [above=of T1]                                   {};
    \draw[-<-]      (T1)    to [out=90,in=-90]          (up1);
\end{tikzpicture}
\sim\quad\e^{\g_i}_{a_i}(p_i) \,,\qquad
\begin{tikzpicture}[baseline=-0.65ex]
    \node[fTSnode]   (T2)    [label= 70:$p_j$,
                            label= 110:$a_j$,
                            label={[label distance=0.4cm] 93:$\g_j$}]       {};
    \node           (up2)   [above=of T2]                                   {};
    \draw[->-]      (T2)    to [out=90,in=-90]          (up2);
\end{tikzpicture}
\sim\quad\e^{\g_j*}_{a_j}(p_j) \,.
\end{equation}
Since we will sum over the spin indices, each line yields a term of $ (-\w^{\g_i\g_j}+ p_i^{\g_i} p_j^{\g_j}/m^2) $.

Each operator ``$ T $" in the red box produces a $ t\,^{\m_l\n_l}_{(l)~\g_i\g_j}(p_i,p_j) $
\begin{equation}
\begin{tikzpicture}[baseline=-0.65ex]
    \node[TSnode]   (T1)    [label=70:$ p_i $,
                            label=-70:$ p_j $,
                            label=110:$ a_i $,
                            label=-110:$ a_j $,
                            label={[label distance=0.4cm] 93:$\g_i$},
                            label={[label distance=0.4cm]-93:$\g_j$}]                     {$ T_{(l)}^{\m_l\n_l}$};
    \node           (up)    [above=of T1]{};
    \node           (down)  [below=of T1]{};
    \draw[-<-]      (T1)    to [out=90,in=-90]          (up);
    \draw[-<-]      (T1)    to [out=-90,in=90]          (down);
\end{tikzpicture}
\sim\, t\,^{\m_l\n_l}_{(l)~\g_i\g_j} (p_i,p_j)
\end{equation}
with $ (l) $ standing for the serial number of the red box.
If $ p_i $ enters the box and $ p_j $ exits the box, this should yield $ t~^{\m_l\n_l}_{(l)~\g_i\g_j} (p_i,-p_j) $, which resembles the vortex rule in the standard Feynman rules.

Based on the above rules, we have included most of the terms in \eq{example:TT}, except for the hydrodynamic parameters, $ \pt\b $, the integrals over ``$ \l $"s, and the integrals over space and momenta.
Of these, the last is the most complex. Due to space limitations, we present three examples
\begin{equation}
\label{example:D1}
\begin{tikzpicture}[baseline=-0.65ex]
    \node[TSnode]   (T1)    [label=70:$ p_i $,
                            label=-70:$ p_j $,
                            label={ 0:}]                     {};
    \node           (up)    [above=of T1]{};
    \node           (down)  [below=of T1]{};
    \node at (6,0) (eq1) {
        $\begin{aligned}
            & \int_{\Xi} d\Xi_{\m_l}(y_l)\int\idp{3}{\bp_i}\idp{3}{\bp_j} e^{-i\,(p_i+p_j)\cdot (y_l-x)}\\
            =& \int\idp{3}{\bp_i}\idp{3}{\bp_j} (2\p)^3\d^{(3)}(\bp_i+\bp_j) \htt_{\m_l} \,,
        \end{aligned}$};
    \draw[-<-]      (T1)    to [out=90,in=-90]          (up);
    \draw[-<-]      (T1)    to [out=-90,in=90]          (down);
\end{tikzpicture}
\end{equation}
\begin{equation}
\label{example:D2}
\begin{tikzpicture}[baseline=-0.65ex]
    \node[TSnode]   (T1)    [label=70:$ p_i $,
                            label=-70:$ p_j $,
                            label={ 0:}]                     {$ D_{(l)}^{\a_m} $};
    \node           (up)    [above=of T1]{};
    \node           (down)  [below=of T1]{};
    \node at (7,0) (eq1) {
        $\begin{aligned}
            & \int_{\Xi} d\Xi_{\m_l}(y_l) (y_l-x)^{\a_m} e^{-i\,(p_i+p_j)\cdot (y_l-x)}\\
            \sim~ & (2\p)^3\d^{(3)}(\bp_i+\bp_j) \htt_{\m_l} \lb-\frac{i}{2}\rb \lb \na_{i}^{\a_m} +\na_{j}^{\a_m} \rb \,,
        \end{aligned}$
    };
    \draw[-<-]      (T1)    to [out=90,in=-90]          (up);
    \draw[-<-]      (T1)    to [out=-90,in=90]          (down);
\end{tikzpicture}
\end{equation}
\begin{equation}
\label{example:D3}
\begin{tikzpicture}[baseline=-0.65ex]
    \node[TSnode]   (T1)    [label=70:$ p_i $,
                            label=-70:$ p_j $,
                            label={ 0:}]                     {$ D_{(l)}^{\a_m} $};
    \node           (up)    [above=of T1]{};
    \node           (down)  [below=of T1]{};
    \node at (7,0) (eq1) {
        $\begin{aligned}
            & \int_{\Xi} d\Xi_{\m_l}(y_l) (y_l-x)^{\a_m} e^{-i\,(p_i-p_j)\cdot (y_l-x)}\\
            \sim~ & (2\p)^3\d^{(3)}(\bp_i-\bp_j) \htt_{\m_l} \lb-\frac{i}{2}\rb \lb \na_{i}^{\a_m} -\na_{j}^{\a_m} \rb
        \end{aligned}$
    };
    \draw[-<-]      (T1)    to [out=90,in=-90]          (up);
    \draw[->-]      (T1)    to [out=-90,in=90]          (down);
\end{tikzpicture}
\end{equation}
with ``$ \sim $ " denoting that the calculation is done with the integrals over the momenta, $ N $ the total number of the red boxes in the diagram,
$ i $ taking a value from $ 1 $ to $ 2N-1 $ ($ j=i+1 $) and $ m $ in the $ \a_m $ standing for the serial number of the ``$ D $".
To verify these equations, we recommend transforming the variables by $ l_1=p_i-p_j,\,l_2=(p_i+p_j)/2 $,
otherwise swapping the delta function and the partial derivatives may be difficult.
Another technique used here is that we multiply an extra exponential term
$
    e^{-i(p_0 \pm p_1 \pm \cdots - p_{(2N+1)})\cdot x} \,,
$
which always equals $ 1 $ by the restriction from the delta functions on the lines.
For example, in \eq{example:TT}, the extra term is $ e^{-i(p_0 - p_1 - p_2 + p_3 + p_4 - p_5)\cdot x} $.
With this term, the $ x $ dependence, except for the hydrodynamic fields, \eg, $ \pt_\a\b_\n (x)$, always appears in $ (y_l-x) $.
Therefore, we can deal with $ x $ together with $ y_l $. 

Based on those examples, a delta function emerges from each red box,
which constrains the momenta across the red box to be the same (opposite) if the line preserves (reverses) its direction across the box.
Therefore, we denote
\begin{equation*}
\begin{tikzpicture}[baseline=-0.65ex]
    \node[TSnode]   (T1)    [label=70:$ p_i $,
                            label=-70:$ p_j $,
                            label={ 0: $\sim$}]                     {$ D_{(l)}^{\a_m} $};
    \node           (up)    [above=of T1]{};
    \node           (down)  [below=of T1]{};
    \draw[-<-]      (T1)    to [out=90,in=-90]          (up);
    \draw[-<-]      (T1)    to [out=-90,in=90]          (down);
\end{tikzpicture}
 D_{(l)}^{\a_m} = \lb-\frac{i}{2}\rb \lb\na_{i}^{\a_m} +\na_{j}^{\a_m}\rb \,.
\end{equation*}
These ``$ D $"s (whether from the same box or not) commute with each other and commute with the delta functions from other boxes.
Therefore, we can apply this rule to diagrams with multiple boxes and multiple ``$ D $"s, without worrying about the sequence of ``$ D $"s.
We should note that here $ \na_{i}^{\a_m} $ denotes the transverse momentum derivative with respect to $p_i$,
with the time component $\na_i^0=0$ by its definition.

However, we note that ``$ D $"s will act on the delta functions from the lines and the $ \d^{(4)}(k-p_0/2-p_{(2N+1)}/2) $ from the MVSD.
Therefore, the calculation is still not trivial. We need to swap ``$ D $"s with these delta functions. (For details, see appendix~\ref{sec:D}.)
These swaps convert ``$ D $"s into
\begin{equation}
    D^{\a_m}_{(l)} = \lb -\frac{i}{2}\rb \sum_{i=0}^{2N+1}\s^{(l)}_i \na^{\a_m}_i \,,
\end{equation}
where $ \sum_{i} $ denotes summing over all the ``$ p $"s (in our case, from $ p_0 $ to $ p_5 $)
and $ \s^{(l)}_i = \pm1 $  ($ \s_i^{(l)} = 1 $ if the line of $ p_i $ points to the box $ l $, otherwise $ \s_i^{(l)} = -1 $).
We should point out that we have neglected the derivatives acting on $ \d(k^0 -E_{\bp_0}/2-E_{\bp_{(2N+1)}}/2) $,
indicating that we have neglected the off-shell terms.
After moving all the delta functions, from boxes and lines, to the left-hand side of the derivatives,
we can integrate them out (except for $ \d(k^0 -E_{\bp_0}/2-E_{\bp_{(2N+1)}}/2) $ ).
This restricts $p_i$ to either $k$ or $\overline{k}$, where $ k^\m=(E_\bk,\bk) $ and $ \overline{k}^\m=(E_\bk,-\bk) $. Due to the delta functions, We need to assign $ p_i=k $ if the line of $ p_i $ points to the bottom of the green box, and assign $ p_i=\overline{k} $ otherwise.

In conclusion, the Feynman rules are as follows.

\begin{tikzpicture}
    \node           at (0,0)            (step1)         [label={0: Write out $\opt{\mathcal{B}}_n$ and draw all the diagrams by utilizing the Wick Theorem.}]        {1)};
\end{tikzpicture}

\begin{tikzpicture}
    \node           at (0,2)            (step1)         [label={0:MVSD }]        {2)};
    \node[fnode]    at (3,1)            (f)             [label={0: $ \quad\sim\quad \d(k^2-m^2) \h(k^0) \frac{2E_\bk}{2E_{\bp_1}2E_{\bp_{(2N)}}} \e^{\g_0}_s(k)\e^{\g_{(2N+1)}*}_r(k) $}]                                                  {$f$};
    \node           at (0,0)            (step2)         [label={0: Lines}]                                                            {3)};
    \node[]         at (3,-1.5)         (O1)            [label={0:$\quad\sim\quad 2 (\htt\cdot p_i) n_B(\b\cdot p_i)(-\w^{\g_i\g_j}+p_i^{\g_i}p_i^{\g_j}/m^2)$},
                                                         label={ -160: $p_j$},
                                                         label={ 160: $\g_j$}]                                                                              {};
    \node[]         at (1,-1.5)         (lftO1)         [label={ 20: $\g_i$},
                                                         label={ -20: $p_i$}]                                                                              {};
    \node[]         at (3,-3)           (O2)            [label={0:$\quad\sim\quad 2 (\htt\cdot p_i) \ls n_B(\b\cdot p_i)+1\rs(-\w^{\g_i\g_j}+p_i^{\g_i}p_i^{\g_j}/m^2)$},
                                                         label={ -160: $p_j$},
                                                         label={ 160: $\g_j$}]                                                                              {};
    \node[]         at (1,-3)           (lftO2)         [label={ -20: $p_i$},
                                                         label={ 20: $\g_i$}]                                                                              {};
    \node           at (1,-4.1)           (text1)         [label={0:The momenta connected by the same line should be equal, $p_i=p_j$.}]           {};
    \draw[-<-]      (O1)    to [out=0,in=180]           (lftO1);
    \draw[->-]      (O2)    to [out=0,in=180]           (lftO2);
\end{tikzpicture}

\begin{tikzpicture}
    \node           at (0,0)            (step1)         [label={0: Vertices }]            {4)};
    \node[TSnode]   at (2,-2)           (T)             [label={60: $p_i$},
                                                         label={120: $\g_i$},
                                                         label={-60: $p_j$},
                                                         label={-120: $\g_j$}]          {$T^{\m_l\n_l}_{(l)}$};
    \node           at (2.3,-2.6)       (eq1)           [label={0:
                                                            \small{$\begin{aligned}
                                                                \quad\sim\quad t\,^{\m_l\n_l}_{(l)~~\g_i\g_j}&(p_i,p_j) = \frac{e^{-\l_l \b\cdot(p_i +p_j)}}{2E_{\bp_i}2E_{\bp_j}} \Big[p_i^{\m_l}p_j^{\n_l}\w_{\g_i\g_j} -p_{i,\g_j}p_j^{\n_l}\w^{\m_l}_{~~\g_i}\\
                                                                &-\frac{1}{2}(p_i\cdot p_j)\w^{\m_l\n_l}\w_{\g_i\g_j} +\frac{1}{2}p_{i,\g_j}p_{j,\g_i}\w^{\m_l\n_l} -\frac{1}{2}m^2\w^{\m_l\n_l}\w_{\g_i\g_j}\Big]
                                                            \end{aligned}$}}]            {};
    \node                               (upT)           [above=of T]                    {};
    \node                               (dwT)           [below=of T]                    {};
    \node[TSnode]   at (2,-5)           (S)             [label={60: $p_i$},
                                                         label={120: $\g_i$},
                                                         label={-60: $p_j$},
                                                         label={-120: $\g_j$}]          {$S^{\m_l\r_l\s_l}_{(l)}$};
    \node           at (2.3,-5.2)       (eq2)           [label={0:
                                                            \small{$\begin{aligned}
                                                                \quad\sim\quad s\,^{\m_l\r_l\s_l}_{(l)~~~~\g_i\g_j}&(p_i,p_j) = 2i\,\frac{e^{-\l_l \b\cdot(p_i +p_j)}}{2E_{\bp_i}2E_{\bp_j}} \Big(p_i^{\m_l}\w^{\r_l}_{~~\g_i}\w^{\s_l}_{~~\g_j}-p_i^{\r_l}\w^{\m_l}_{~~\g_i}\w^{\s_l}_{~~\g_j}\Big)
                                                            \end{aligned}$}}]            {};
    \node                               (upS)           [above=of S]                    {};
    \node                               (dwS)           [below=of S]                    {};
    \draw[-<-]      (T)     to [out=90, in=-90]         (upT);
    \draw[-<-]      (T)     to [out=-90, in=90]         (dwT);
    \draw[-<-]      (S)     to [out=90, in=-90]         (upS);
    \draw[-<-]      (S)     to [out=-90, in=90]         (dwS);
\end{tikzpicture}

\begin{tikzpicture}
    \node           at (0,0)            (step1)         [label={0: Derivatives }]            {5)};
    \node[TSnode]   at (2,-2)           (T)             [label={60: $p_i$},
                                                         label={120: $\g_i$},
                                                         label={-60: $p_j$},
                                                         label={-120: $\g_j$}]          {$D^{\a_m}_{(l)}$};
    \node           at (2.3,-2.1)       (eq1)           [label={0:
                                                            $\begin{aligned}
                                                                \quad\sim\quad D^{\a_m}_{(l)} = \lb -\frac{i}{2}\rb \sum_{i=0}^{2N+1}\s^{(l)}_i \na^{\a_m}_i
                                                            \end{aligned}$}]            {};
    \node                               (upT)           [above=of T]                    {};
    \node                               (dwT)           [below=of T]                    {};
    \node           at (1,-4)           (text1)         [label={0:If the line of $ p_i $ pointing to the box $ l $
                                                        then $ \s_i^{(l)} = 1 $, otherwise $ \s_i^{(l)} = -1 $.}]           {};
     \node           at (1,-4.7)           (text1)         [label={0:It should be put on the left-hand side of all lines and                                                                          vertices.}]           {};
    \draw[-<-]      (T)     to [out=90, in=-90]         (upT);
    \draw[-<-]      (T)     to [out=-90, in=90]         (dwT);
\end{tikzpicture}

\begin{tikzpicture}
    \node           at (0,0)            (step1)         [label={0: Momentum conservation}]                                  {6)};
    \node           at (1,-1)           (text1)         [label={0: After evaluating the ``$D$" derivatives, replace  $ p_i$ by $k$ if the line}]  {};
    \node           at (1,-1.7)           (text1)         [label={0: of $ p_i $ points to the bottom end of the green box, otherwise $ p_i=\overline{k} $.}]  {};
\end{tikzpicture}

\begin{tikzpicture}
    \node           at (0,0)            (step1)         [label={0: Integrate over all the $\l$}]                                  {7)};
    \node           at (1,-1)           (text1)         [label={0: $\int^1_0 d\l_1 \int^{\l_1}_0 d\l_2 \cdots \int^{\l_{n-1}}_0 d\l_n$}]  {};
\end{tikzpicture}

For example,
\small{\begin{equation}
\label{example:2TT}
\begin{split}
&
\begin{tikzpicture}
    \node[TSnode]   (T1)    [label={[label distance=0.1cm] 70:$p_1$},
                            label={[label distance=0.1cm]-70:$p_2$},
                            label={[label distance=0.1cm] 110:$a_1$},
                            label={[label distance=0.1cm]-110:$a_2$},
                            label={[label distance=0.3cm] 90:$\g_1$},
                            label={[label distance=0.3cm]-90:$\g_2$}]                   {$D_{(1)}^{\a_1}T_{(1)}^{\m_1\n_1}$};
    \node[TSnode]   (T2)    [right=of T1,
                            label={[label distance=0.1cm] 70:$p_3$},
                            label={[label distance=0.1cm]-70:$p_4$},
                            label={[label distance=0.1cm] 110:$a_3$},
                            label={[label distance=0.1cm]-110:$a_4$},
                            label={[label distance=0.3cm] 90:$\g_3$},
                            label={[label distance=0.3cm]-90:$\g_4$}]                   {$D_{(2)}^{\a_2}T_{(2)}^{\m_2\n_2}$};
    \node[fnode]    (f)     [right=of T2,
                            label=70:\small{$ p_0 $},
                            label=-70:\small{$ p_5 $},
                            label={[label distance=0.3cm] 90:$\g_0$},
                            label={[label distance=0.3cm]-90:$\g_5$},
                            label=110:$ a_0 $, label=-110:$ a_5 $]      {$f$};
    \draw[->-]  (f.north)     to [out=90, in=90]      (T1.north);
    \draw[-<-]  (T1.south)    to [out=-90, in=90]     (T2.north);
    \draw[->-]  (T2.south)    to [out=-90, in=-90]    (f.south);
\end{tikzpicture}\\
=& \d(k^2-m^2) \e^{\g_0}_s(k)\e^{\g_{3*}}_r(k) \int_0^1 d\l_1 \int_0^{\l_1}d\l_2\, \Bigg\{D^{\a_1}_{(1)}
\, D^{\a_2}_{(2)}\, \frac{2E_\bk}{2E_{\bp_1}2E_{\bp_{(2N)}}} t^{\m_1\n_1}_{(1)~\g_1\g_2}(p_1,p_2)\\
&\times t^{\m_2\n_2}_{(2)~\g_3\g_4}(-p_3,-p_4)(2\htt\cdot p_1)[n_B(\b\cdot p_1)+1)](2\htt\cdot p_2)[n_B(\b\cdot p_2)+1](2\htt\cdot p_4)n_B(\b\cdot p_4)\\
&\times \lb-\w^{\g_0\g_1}+\frac{p_1^{\g_0}p_1^{\g_1}}{m^2}\rb \lb-\w^{\g_2\g_3}+\frac{p_2^{\g_2}p_2^{\g_3}}{m^2}\rb \lb-\w^{\g_4\g_5}+\frac{p_4^{\g_4}p_4^{\g_5}}{m^2}\rb \Bigg\}\bigg|^{p_1=k,p_2=\overline{k}}_{p_3=\overline{k},p_4=k}
\end{split}
\end{equation}
}
with
\begin{equation}
    D^{\a_1}_{(1)} = \lb -\frac{i}{2}\rb \lb\w^{\a_1\z_1}-\hat{t}^{\a_1}\htt^{\z_1}\rb \lb\frac{\pt}{\pt p_{1}^{\z_1}} +\frac{\pt}{\pt p_{2}^{\z_1}}
    +\frac{\pt}{\pt p_{3}^{\z_1}} -\frac{\pt}{\pt p_{4}^{\z_1}} \rb
\end{equation}
and
\begin{equation}
    \begin{split}
    D^{\a_2}_{(2)} &= \lb -\frac{i}{2}\rb \lb\w^{\a_2\z_2}-\hat{t}^{\a_2}\htt^{\z_2}\rb \lb\frac{\pt}{\pt p_{1}^{\z_2}} -\frac{\pt}{\pt p_{2}^{\z_2}} -\frac{\pt}{\pt p_{3}^{\z_2}} -\frac{\pt}{\pt p_{4}^{\z_2}}\rb.
    \end{split}
\end{equation}

So far, we have set all momenta to be on the normal mass shell $ p_i^{\m} = (E_{\bp_i},\bp_i) $.
In practice, we can relax these restrictions by replacing $E_{\bp_i}$ by an energy $p_i^0$,
which is treated as being independent of $\bp_i$ during the calculation, for Feynman rules of the MVSD, lines, and vertices.
The related ``$D$"s should be modified to the following form
\begin{equation}
     D^{\a_m}_{(l)} = \lb-\frac{i}{2}\rb \sum_{i=0}^{2N+1}\s^{(l)}_i\ls \pt^{\a_m}_i-\hat{p}_i^{\a_m}\lb\htt\cdot\pt_i\rb \rs \,.
\end{equation}
After evaluating the ``$D$" derivatives, we still need to assign  $ p_i=k $ or $\overline{k}$,
with $ k^\m=(E_\bk,\bk) $ and $ \overline{k}^\m=(E_\bk,-\bk) $ on the mass shell. 

\section{Second-order results}
\label{sec:2nd:order}
Using the diagrammatic scheme discussed in the previous section, the six terms for the second-order MVSD in \eq{expand:f2}  can be expressed as follows.
\begin{equation}
\begin{split}
\label{MVSD:TT}
f^{(2)}_{rs}|_{TT} &= \htt_{\m_1}\htt_{\m_2}[\pt_{\a_1}\b_{\n_1}](x)[\pt_{\a_2}\b_{\n_2}](x)\\
&\times\bigg\{
\begin{tikzpicture}[baseline=-0.65ex]
    \node[TSnode]   at (0,0)            (T1)            [anchor=west]                   {$D_{(1)}^{\a_1}\,T^{\m_1\n_1}_{(1)}$};
    \node[TSnode]   at (2.1,0)          (T2)            [anchor=west]                   {$D_{(2)}^{\a_2}\,T^{\m_2\n_2}_{(2)}$};
    \node[fnode]    at (4.5,0)          (f1)            [anchor=west,
                                                        label={150: $s$},
                                                        label={-150: $r$}]              {$f$};
    \draw[->-]      (f1.north)      to [out=120,in=-60]                             (T1.south);
    \draw[->-]      (T1.north)      to [out=60,in=120]                              (T2.north);
    \draw[->-]      (T2.south)      to [out=-60,in=-120]                            (f1.south);
\end{tikzpicture}
+\mathrm{other\ 15\ diagrams}\bigg\} \,,
\end{split}
\end{equation}
\begin{equation}
\begin{split}
f^{(2)}_{rs}|_{TS} &= -\frac{1}{2}\htt_{\m_1}\htt_{\m_2}[\pt_{\a_1}\b_{\n_1}](x)\O_{\r_2\s_2}(x)\\
&\times\bigg\{
\begin{tikzpicture}[baseline=-0.65ex]
    \node[TSnode]   at (0,0)            (T1)            [anchor=west]                   {$D_{(1)}^{\a_1}\,T^{\m_1\n_1}_{(1)}$};
    \node[TSnode]   at (2.4,0)          (T2)            [anchor=west]                   {$S^{\m_2\r_2\s_2}_{(2)}$};
    \node[fnode]    at (4.5,0)          (f1)            [anchor=west,
                                                        label={150: $s$},
                                                        label={-150: $r$}]              {$f$};
    \draw[->-]      (f1.north)      to [out=120,in=-60]                             (T1.south);
    \draw[->-]      (T1.north)      to [out=60,in=120]                              (T2.north);
    \draw[->-]      (T2.south)      to [out=-60,in=-120]                            (f1.south);
\end{tikzpicture}
+\mathrm{other\ 15\ diagrams}\bigg\} \,,
\end{split}
\end{equation}
\begin{equation}
\begin{split}
f^{(2)}_{rs}|_{ST} &= -\frac{1}{2}\htt_{\m_1}\htt_{\m_2}\O_{\r_1\s_1}(x)[\pt_{\a_2}\b_{\n_2}](x)\\
&\times\bigg\{
\begin{tikzpicture}[baseline=-0.65ex]
    \node[TSnode]   at (0,0)            (T1)            [anchor=west]                   {$S^{\m_1\r_1\s_1}_{(1)}$};
    \node[TSnode]   at (1.8,0)          (T2)            [anchor=west]                   {$D_{(2)}^{\a_2}\,T^{\m_2\n_2}_{(2)}$};
    \node[fnode]    at (4.5,0)          (f1)            [anchor=west,
                                                        label={150: $s$},
                                                        label={-150: $r$}]              {$f$};
    \draw[->-]      (f1.north)      to [out=120,in=-60]                             (T1.south);
    \draw[->-]      (T1.north)      to [out=60,in=120]                              (T2.north);
    \draw[->-]      (T2.south)      to [out=-60,in=-120]                            (f1.south);
\end{tikzpicture}
+\mathrm{other\ 15\ diagrams}\bigg\} \,,
\end{split}
\end{equation}
\begin{equation}
\begin{split}
\label{MVSD:SS}
f^{(2)}_{rs}|_{SS} &= \frac{1}{4}\htt_{\m_1}\htt_{\m_2}\O_{\r_1\s_1}(x)\O_{\r_2\s_2}(x)\\
&\times\bigg\{
\begin{tikzpicture}[baseline=-0.65ex]
    \node[TSnode]   at (0,0)            (T1)            [anchor=west]                   {$S^{\m_1\r_1\s_1}_{(1)}$};
    \node[TSnode]   at (2.2,0)          (T2)            [anchor=west]                   {$S^{\m_2\r_2\s_2}_{(2)}$};
    \node[fnode]    at (4.5,0)          (f1)            [anchor=west,
                                                        label={150: $s$},
                                                        label={-150: $r$}]              {$f$};
    \draw[->-]      (f1.north)      to [out=120,in=-60]                             (T1.south);
    \draw[->-]      (T1.north)      to [out=60,in=120]                              (T2.north);
    \draw[->-]      (T2.south)      to [out=-60,in=-120]                            (f1.south);
\end{tikzpicture}
+\mathrm{other\ 15\ diagrams}\bigg\} \,,
\end{split}
\end{equation}
\begin{equation}
\begin{split}
\label{MVSD:2T}
f^{(2)}_{rs}|_{T} &= -\htt_{\m_1}[\pt_{\a_1}\pt_{\a_2}\b_{\n_1}](x)\\
&\times\bigg\{
\begin{tikzpicture}[baseline=-0.65ex]
    \node[TSnode]   at (0,0)            (T1)            [anchor=west]                   {$D^{\a_1}_{(1)}D^{\a_2}_{(1)}T^{\m_1\n_1}_{(1)}$};
    \node[fnode]    at (3,0)            (f1)            [anchor=west,
                                                        label={150: $s$},
                                                        label={-150: $r$}]              {$f$};
    \draw[->-]      (f1.north)      to [out=120,in=60]                              (T1.north);
    \draw[->-]      (T1.south)      to [out=-60,in=-120]                            (f1.south);
\end{tikzpicture}
+
\begin{tikzpicture}[baseline=-0.65ex]
    \node[TSnode]   at (0,0)            (T1)            [anchor=west]                   {$D^{\a_1}_{(1)}D^{\a_2}_{(1)}T^{\m_1\n_1}_{(1)}$};
    \node[fnode]    at (3,0)            (f1)            [anchor=west,
                                                        label={150: $s$},
                                                        label={-150: $r$}]              {$f$};
    \draw[->-]      (f1.north)      to [out=110,in=-70]                             (T1.south);
    \draw[->-]      (T1.north)      to [out=70,in=-110]                             (f1.south);
\end{tikzpicture}
\bigg\} \,,
\end{split}
\end{equation}
\begin{equation}
\begin{split}
\label{MVSD:2S}
f^{(2)}_{rs}|_{S} &= \frac{1}{2}\htt_{\m_1}[\pt_{\a_1}\O_{\r_1\s_1}](x)\\
&\times\bigg\{
\begin{tikzpicture}[baseline=-0.65ex]
    \node[TSnode]   at (0,0)            (T1)            [anchor=west]                   {$D^{\a_1}_{(1)}S^{\m_1\r_1\s_1}_{(1)}$};
    \node[fnode]    at (3,0)            (f1)            [anchor=west,
                                                        label={150: $s$},
                                                        label={-150: $r$}]              {$f$};
    \draw[->-]      (f1.north)      to [out=120,in=60]                              (T1.north);
    \draw[->-]      (T1.south)      to [out=-60,in=-120]                            (f1.south);
\end{tikzpicture}
+
\begin{tikzpicture}[baseline=-0.65ex]
    \node[TSnode]   at (0,0)            (T1)            [anchor=west]                   {$D^{\a_1}_{(1)}S^{\m_1\r_1\s_1}_{(1)}$};
    \node[fnode]    at (3,0)            (f1)            [anchor=west,
                                                        label={150: $s$},
                                                        label={-150: $r$}]              {$f$};
    \draw[->-]      (f1.north)      to [out=110,in=-70]                             (T1.south);
    \draw[->-]      (T1.north)      to [out=70,in=-110]                             (f1.south);
\end{tikzpicture}
\bigg\} \,.
\end{split}
\end{equation}
Note that for $ f^{(2)}_{rs}|_{TT},\,f^{(2)}_{rs}|_{TS},\,f^{(2)}_{rs}|_{ST},\,f^{(2)}_{rs}|_{SS} $, they contain 16 diagrams each

{\centering
\begin{tikzpicture}
    \node[TSnode]   at(0,0)             (T1)            []                  {};
    \node[TSnode]   at(1,0)             (T2)            []                  {};
    \node[fnode]    at(2,0)             (f1)            []                  {};
    \node[TSnode]   at(3,0)             (T3)            []                  {};
    \node[TSnode]   at(4,0)             (T4)            []                  {};
    \node[fnode]    at(5,0)             (f2)            []                  {};
    \node[TSnode]   at(6,0)             (T5)            []                  {};
    \node[TSnode]   at(7,0)             (T6)            []                  {};
    \node[fnode]    at(8,0)             (f3)            []                  {};
    \node[TSnode]   at(9,0)             (T7)            []                  {};
    \node[TSnode]   at(10,0)            (T8)            []                  {};
    \node[fnode]    at(11,0)            (f4)            []                  {};
    \node[TSnode]   at(0,-1.5)          (T9)            []                  {};
    \node[TSnode]   at(1,-1.5)          (T10)           []                  {};
    \node[fnode]    at(2,-1.5)          (f5)            []                  {};
    \node[TSnode]   at(3,-1.5)          (T11)           []                  {};
    \node[TSnode]   at(4,-1.5)          (T12)           []                  {};
    \node[fnode]    at(5,-1.5)          (f6)            []                  {};
    \node[TSnode]   at(6,-1.5)          (T13)           []                  {};
    \node[TSnode]   at(7,-1.5)          (T14)           []                  {};
    \node[fnode]    at(8,-1.5)          (f7)            []                  {};
    \node[TSnode]   at(9,-1.5)          (T15)           []                  {};
    \node[TSnode]   at(10,-1.5)         (T16)           []                  {};
    \node[fnode]    at(11,-1.5)         (f8)            []                  {};
    \node[TSnode]   at(9,-1.5)          (T15)           []                  {};
    \node[TSnode]   at(10,-1.5)         (T16)           []                  {};
    \node[fnode]    at(11,-1.5)         (f8)            []                  {};
    \node[TSnode]   at(0,-3)            (T17)           []                  {};
    \node[TSnode]   at(1,-3)            (T18)           []                  {};
    \node[fnode]    at(2,-3)            (f9)            []                  {};
    \node[TSnode]   at(3,-3)            (T19)           []                  {};
    \node[TSnode]   at(4,-3)            (T20)           []                  {};
    \node[fnode]    at(5,-3)            (f10)           []                  {};
    \node[TSnode]   at(6,-3)            (T21)           []                  {};
    \node[TSnode]   at(7,-3)            (T22)           []                  {};
    \node[fnode]    at(8,-3)            (f11)           []                  {};
    \node[TSnode]   at(9,-3)            (T23)           []                  {};
    \node[TSnode]   at(10,-3)           (T24)           []                  {};
    \node[fnode]    at(11,-3)           (f12)           []                  {};
    \node[TSnode]   at(0,-4.5)          (T25)           []                  {};
    \node[TSnode]   at(1,-4.5)          (T26)           []                  {};
    \node[fnode]    at(2,-4.5)          (f13)           []                  {};
    \node[TSnode]   at(3,-4.5)          (T27)           []                  {};
    \node[TSnode]   at(4,-4.5)          (T28)           []                  {};
    \node[fnode]    at(5,-4.5)          (f14)           []                  {};
    \node[TSnode]   at(3,-4.5)          (T27)           []                  {};
    \node[TSnode]   at(4,-4.5)          (T28)           []                  {};
    \node[fnode]    at(5,-4.5)          (f14)           []                  {};
    \node[TSnode]   at(6,-4.5)          (T29)           []                  {};
    \node[TSnode]   at(7,-4.5)          (T30)           []                  {};
    \node[fnode]    at(8,-4.5)          (f15)           []                  {};
    \node[TSnode]   at(9,-4.5)          (T31)           []                  {};
    \node[TSnode]   at(10,-4.5)         (T32)           []                  {};
    \node[fnode]    at(11,-4.5)         (f16)           []                  {};

    \draw[->-]  (f1.north)      to [out=100, in=80]     (T1.north);
    \draw[->-]  (T1.south)      to [out=-80, in=100]    (T2.north);
    \draw[->-]  (T2.south)      to [out=-80, in=-100]    (f1.south);
    \draw[->-]  (f2.north)      to [out=100, in=80]     (T3.north);
    \draw[-<-]  (T3.south)      to [out=-80, in=100]    (T4.north);
    \draw[->-]  (T4.south)      to [out=-80, in=-100]   (f2.south);
    \draw[->-]  (f3.north)      to [out=100, in=80]     (T5.north);
    \draw[->-]  (T5.south)      to [out=-80, in=-100]   (T6.south);
    \draw[->-]  (T6.north)      to [out=80, in=-100]    (f3.south);
    \draw[->-]  (f4.north)      to [out=100, in=80]     (T7.north);
    \draw[-<-]  (T7.south)      to [out=-80, in=-100]   (T8.south);
    \draw[->-]  (T8.north)      to [out=80, in=-100]    (f4.south);
    \draw[->-]  (f5.north)      to [out=100, in=-80]    (T9.south);
    \draw[->-]  (T9.north)      to [out=80, in=100]     (T10.north);
    \draw[->-]  (T10.south)     to [out=-80, in=-100]   (f5.south);
    \draw[->-]  (f6.north)      to [out=100, in=-80]    (T11.south);
    \draw[-<-]  (T11.north)     to [out=80, in=100]     (T12.north);
    \draw[->-]  (T12.south)     to [out=-80, in=-100]   (f6.south);
    \draw[->-]  (f7.north)      .. controls +(up:15mm) and +(down:10mm) ..      (T13.south);
    \draw[->-]  (T13.north)     .. controls +(up:10mm) and +(down:10mm) ..      (T14.south);
    \draw[->-]  (T14.north)     .. controls +(up:10mm) and +(down:10mm) ..      (f7.south);
    \draw[->-]  (f8.north)      .. controls +(up:15mm) and +(down:10mm) ..      (T15.south);
    \draw[-<-]  (T15.north)     .. controls +(up:10mm) and +(down:10mm) ..      (T16.south);
    \draw[->-]  (T16.north)     .. controls +(up:10mm) and +(down:10mm) ..      (f8.south);
    \draw[->-]  (f9.north)      to [out=100, in=80]     (T18.north);
    \draw[->-]  (T18.south)     to [out=-100, in=-80]   (T17.south);
    \draw[->-]  (T17.north)     to [out=80, in=-100]    (f9.south);
    \draw[->-]  (f10.north)     to [out=100, in=80]     (T20.north);
    \draw[-<-]  (T20.south)     to [out=-100, in=-80]   (T19.south);
    \draw[->-]  (T19.north)     to [out=80, in=-100]    (f10.south);
    \draw[->-]  (f11.north)     .. controls +(up:10mm) and +(down:10mm) ..      (T22.south);
    \draw[->-]  (T22.north)     .. controls +(up:10mm) and +(down:10mm) ..      (T21.south);
    \draw[->-]  (T21.north)     .. controls +(up:10mm) and +(down:15mm) ..      (f11.south);
    \draw[->-]  (f12.north)     .. controls +(up:10mm) and +(down:10mm) ..      (T24.south);
    \draw[-<-]  (T24.north)     .. controls +(up:10mm) and +(down:10mm) ..      (T23.south);
    \draw[->-]  (T23.north)     .. controls +(up:10mm) and +(down:15mm) ..      (f12.south);
    \draw[->-]  (f13.north)     to [out=100, in=80]     (T26.north);
    \draw[->-]  (T26.south)     to [out=-100, in=80]    (T25.north);
    \draw[->-]  (T25.south)     to [out=-80, in=-100]   (f13.south);
    \draw[->-]  (f14.north)     to [out=100, in=80]     (T28.north);
    \draw[-<-]  (T28.south)     to [out=-100, in=80]    (T27.north);
    \draw[->-]  (T27.south)     to [out=-80, in=-100]   (f14.south);
    \draw[->-]  (f15.north)     to [out=100, in=-80]    (T30.south);
    \draw[->-]  (T30.north)     to [out=80, in=100]     (T29.north);
    \draw[->-]  (T29.south)     to [out=-80, in=-100]   (f15.south);
    \draw[->-]  (f16.north)     to [out=100, in=-80]    (T32.south);
    \draw[-<-]  (T32.north)     to [out=80, in=100]     (T31.north);
    \draw[->-]  (T31.south)     to [out=-80, in=-100]   (f16.south);
\end{tikzpicture}.\\
}
One can utilize FeynCalc~\cite{Mertig:1990an,Shtabovenko:2016sxi,Shtabovenko:2020gxv} to evaluate the diagrams.
The final results are
\begin{equation}
\label{final:S}
    \begin{split}
    f_{rs}^{(2)}|_S(x,k)
    =& \d(k^2-m^2)\h(k^0) n_B(1+n_B) \e^{r*}_{(\g_3}(k)\e^{s}_{\g_0)}(k) \frac{1}{4E_k}[\pt^{\perp}_{\a_1}\O_{\r_1}^{\ \,\g_0}(x)]\\
    & \times\ls \hat{k}^{\a_1}\htt^{\r_1}{\htt}^{\g_3} -\g_k^2\,\hat{k}^{\r_1}\w^{\g_3\a_1} -\lb\w^{\a_1\r_1} -\frac{k^{\a_1}k^{\r_1}}{m^2} \rb{\htt}{^{\g_3}}\rs \,,
    \end{split}
\end{equation}
\begin{equation}
\label{MVSD:f2T}
\begin{split}
    f_{rs}^{(2)}|_T(x,k) &= \d(k^2-m^2)\h(k^0) n_B(1+n_B) \e^{r*}_{(\g_3}(k)\e^{s}_{\g_0)}(k) \frac{1}{4E_k} [\pt^{\perp}_{\a_1}\pt^{\perp}_{\a_2}\b_{\n_1}(x)]\\
    & \times \Big[ (2\htt^{\n_1} +\gamma_k^2\hat{k}^{\n_1}) \w^{\a_1\g_0}\w^{\g_3\a_2} +( 2\hat{k}^{\a_1}\w^{\a_2\n_1} +\gamma_k^2\hat{k}^{\a_1}\hat{k}^{\a_2}\hat{k}^{\n_1}) \htt^{\g_0}\htt^{\g_3}\\
    & -2(\w^{\a_1\n_1} +\hat{k}^{\a_1}\htt^{\n_1} +\gamma_k^2\hat{k}^{\a_1}\hat{k}^{\n_1}) \w^{\a_2\g_0}\htt^{\g_3}\Big]\\
    & +\d(k^2-m^2)\h(k^0) n_B(1+n_B)\d_{rs}\frac{1}{2E_k} [\pt^{\perp}_{\a_1}\pt^{\perp}_{\a_2}\b_{\n_1}(x)]\\
    & \times \frac{1}{2}\Big\{ \w^{\a_1\a_2}(\htt^{\n_1}+\hat{k}^{\n_1}) +\hat{k}^{\a_1}(2\w^{\a_2\n_1} -\hat{k}^{\a_2}\htt^{\n_1} +3\hat{k}^{\a_2}\hat{k}^{\n_1})\\
    & +(2n_B+1)\gk (\htt\cdot\tilde{\b})(\w^{\a_1\a_2} +\hat{k}^{\a_1}\hat{k}^{\a_2})\hat{k}^{\n_1} +\frac{1}{3}\gk^2[6n_B(1+n_B)+1]\hat{k}^{\n_1}\\
    & \times \ls \tilde{\b}^{\a_1}\tilde{\b}^{\a_2} +(\htt\cdot\tilde{\b})\hat{k}^{\a_1}\hat{k}^{\a_2} -2(\htt\cdot\tilde{\b})\tilde{\b}^{\a_1}\hat{k}^{\a_2}\rs\Big\} \,,
\end{split}
\end{equation}
\begin{align}
\label{final:SS}
\begin{autobreak}
\MoveEqLeft
    f_{rs}^{(2)}|_{SS}(x,k) = 
        \d(k^2-m^2)\h(k^0) n_B(1+n_B) \e^{r*}_{(\g_5}(k)\e^{s}_{\g_0)}(k)\O_{\r_1\s_1}(x)\O_{\r_2\s_2}(x)
        \times \frac{1}{16m^2E_\bk^2}\Bigg\{ (2n_B+1)\bigg[ (E_\bk^2-m^2)k^{\r _1} k^{\r _2}\w^{\g _0 \s _1} \w^{\g _5 \s _2}
        +2m^2 k^{\r _1} k^{\r _2} \htt^{\g _0} \htt^{\s _1} \w^{\g _5 \s _2}
	-m^2 k^{\r _1} k^{\r _2} \htt^{\g _0} \w^{\s _1 \s _2} \htt^{\g _5} 
	-4m^2E_\bk k^{\r _1}\w^{\g _0 \s _2} \htt^{[\g _5} \w^{ \r _2]\s _1}
	-4m^2E_\bk^2 \w^{\r _1 \r _2} \w^{\g _0 \s _1} \w^{\g _5 \s _2}\bigg]
        +A\bigg[-2 E_\bk \bar{k}^{\s _1} \bar{k}^{\s _2} k^{\r _1} \htt^{\g _0} \w^{\g _5 \r _2}
	+( E_\bk^2-m^2 )\bar{k}^{\s _1} \bar{k}^{\s _2}  \w^{\g _5 \r _1} \w^{\g _0 \r _2}
	-m^2\w^{\s _1 \s _2} k^{\r _1} k^{\r _2} \htt^{\g _5} \htt^{\g _0}
	-2m^2 k^{\r _1} \htt^{\g _0} \htt^{\s _1} \bar{k}^{\r _2} \w^{\g _5 \s _2}
	+k^{\r _1} k^{\r _2} \htt^{\g _5} \htt^{\g _0} \bar{k}^{\s _1} \bar{k}^{\s _2}\bigg]\Bigg\} \,,
\end{autobreak}
\end{align}
\begin{align}
\label{final:ST}
\begin{autobreak}
\MoveEqLeft
    f_{rs}^{(2)}|_{TS}(x,k) + f_{rs}^{(2)}|_{ST}(x,k) = 
        \d(k^2-m^2)\h(k^0) n_B(1+n_B) \e^{r*}_{(\g_5}(k)\e^{s}_{\g_0)}(k)[\pt_{\a_1}\b_{\n_1}(x)]\O_{\r_2\s_2}(x)
        \times \frac{1}{8m^2 E_\bk^2}\Bigg\{(2n_B+1)\bigg[-(E_\bk^2+m^2)k^{\n _1} k^{\s _2}\w^{\a _1 \g _5} \w^{\g _0 \r _2}
	+E_\bk k^{\a _1} k^{\n _1} k^{\s _2} \htt^{\g _0} \w^{\g _5 \r _2}
	+2m^2 E_\bk k^{\n _1} \htt^{[\g _0} \w^{\r _2]\a _1 } \w^{\g _5 \s _2}
	+m^2 k^{\n _1} k^{\r _2} \htt^{\g _0} \htt^{\g _5} \w^{\a _1 \s _2}
	-m^2 k^{\n _1} k^{\r _2} \htt^{\g _0} \htt^{(\a _1} \w^{\s _2) \g _5}\bigg]
        +A\bigg[(2E_\bk^2-m^2) \lb\w^{\a _1 \g _0} k^{\r _2} k^{\n _1}-2E_\bk k^{(\a _1}\w^{\r _2)\g _0 } \htt^{\n _1} \rb \htt^{\g _5} \htt^{\s _2}
        +(E_\bk^2-m^2)\bar{k}^{\s _2} \bar{k}^{\n _1} \w^{\a _1 \g _0} \w^{\g _5 \r _2} 
        +2(E_\bk^2-m^2) k^{\n _1} k^{\a _1}  \htt^{\g _0} \htt^{\s _2} \w^{\g _5 \r _2}
	-E_\bk (\w^{\g _0 \s _2}-2\htt^{\g _0} \htt^{\s _2}) \bar{k}^{\n _1} k^{\r _2} \htt^{\g _5} k^{\a _1}
	+m^2 \htt^{[\g _0} \w^{\a _1] \s _2} k^{\r _2} \htt^{\g _5} k^{\n _1}
	+m^2 E_\bk \w^{\a _1 \n _1} (\w^{\g _0 \s _2}-2\htt^{\g _0} \htt^{\s _2})  k^{\r _2} \htt^{\g _5}
	-m^2 E_\bk \w^{\a _1 \n _1} \w^{\g _0 \s _2} \htt^{\g _5} \bar{k}^{\r _2}
	\bigg]\Bigg\} \,,
\end{autobreak}
\end{align}
\begin{align}
\label{final:TT}
\begin{autobreak}
\MoveEqLeft
        f_{rs}^{(2)}|_{TT}(x,k) =  
        \d(k^2-m^2)\h(k^0) n_B(1+n_B) \e^{r*}_{(\g_5}(k)\e^{s}_{\g_0)}(k) [\pt_{\a_1}\b_{\n_1}(x)][\pt_{\a_2}\b_{\n_2}(x)]
        \times\frac{1}{16m^2 E_\bk^2}\Bigg\{(2n_B+1) \bigg[-(E_\bk^2+m^2)k^{\n _2} k^{\n _1}\w^{\a _1 \g _0} \w^{\a _2 \g _5}
	+2E_\bk k^{\a _1} k^{\n _1} k^{\n _2} \htt^{\g _0} \w^{\a _2 \g _5}
	+2m^2 k^{\n _2} k^{\n _1} \htt^{\a _1} \htt^{\g _0} \w^{\a _2 \g _5}
	-m^2\w^{\a _1 \a _2} k^{\n _2} k^{\n _1} \htt^{\g _0} \htt^{\g _5}
	-k^{\a _2} k^{\a _1} k^{\n _1} k^{\n _2} \htt^{\g _0} \htt^{\g _5}\bigg]
        +A\bigg[-4E_\bk(2 E_\bk^2-m^2)\w^{\a _2 \g _0} \htt^{\g _5} \htt^{\n _1} \htt^{\n _2} k^{\a _1}
        +(E_\bk^2-m^2) \w^{\a _1 \g _0} \w^{\a _2 \g _5} \bar{k}^{\n _1} \bar{k}^{\n _2}
        +4(E_\bk^2-m^2) \w^{\a _1 \g _0} k^{\n _2} \htt^{\g _5} \htt^{\n _1} k^{\a _2}  
	+\bar{k}^{\n _1} \bar{k}^{\n _2} \htt^{\g _0} \htt^{\g _5} k^{\a _1} k^{\a _2} 
	+2E_\bk \w^{\a _1 \g _0} \htt^{\g _5} k^{\n _1} \bar{k}^{\n _2} k^{\a _2} 
	+2m^2 \w^{\a _1 \n _1} \htt^{\g _0} \htt^{\g _5} (k^{\n _2}-\bar{k}^{\n _2}) k^{\a _2}
	-m^2\w^{\a _1 \a _2} \htt^{\g _0} \htt^{\g _5} k^{\n _1} k^{\n _2}
        +2m^2\w^{\a _2 \g _0} k^{\n _2} \htt^{\a _1} \htt^{\g _5} k^{\n _1}
        -2m^2 E_\bk \w^{\a _1 \g _0} \w^{\a _2 \n _2} \htt^{\g _5} (k^{\n _1}-\bar{k}^{\n _1})
        \bigg]\Bigg\}
        +\d_{rs}\cdots \,,
\end{autobreak}
\end{align}
where we have introduced the following shorthand notations: $n_B= n_B(\b(x)\cdot k) $, $ \bar{n}_B = n_B(\b\cdot\bar{k})$,
$ \gk =E_{\bk}/m $, $ \pt^{\perp}_{\a_1} =\pt_{\a_1}-\htt_{\a_1}(\htt\cdot\pt)$, $ \tilde{\b} =m \b$, and 
\begin{equation}
    A = \frac{1}{2 \lb \htt \cdot  \b \rb^2 E_\bk^2}\left[2(2\bar{n}_B+1) \sinh^2\lb (\htt\cdot\b)E_\bk \rb -\sinh\lb 2( \htt\cdot\b)E_\bk\rb +2(\htt\cdot\b)E_\bk\right] \,.
\end{equation}
Note that the terms proportional to $ \d_{rs} $ do not contribute to tensor (vector) polarization up to $ \mathcal{O}(\pt^3)$ ($ \mathcal{O}(\pt^2)$),
and hence we do not show such terms in $ f^{(2)}_{rs}|_{TT} $.
With these results, the full second-order expression for the spin density matrix in phase space is obtained through
\begin{equation}
    \H^{(2)}_{rs}(x,k) =\frac{f^{(2)}_{rs}(x,k)}{ f^{(0)}(x,k)}-\frac{f^{(0)}_{rs}(x,k)f^{(2)}(x,k)}{ [f^{(0)}(x,k)]^2} \,,
\end{equation}
where the second term does not contribute to the second-order tensor spin polarization (see \eq{leading:T}). The corresponding Cooper-Frye type formulas for spin density matrix and for vector and tensor spin polarization are obtained through \eqs{eq:cooperfrye}{eq:cooperfrye3}.

We note that the terms proportional to $A$ in eqs.~(\ref{final:SS})-(\ref{final:TT}) may have significant contributions to the spin alignment of vector bosons. For instance, for a thermal velocity $ \b_\m = \hat{t}_\m /T $, $ A $ reduces to
\begin{equation}
    A = \frac{\sinh \lb E_\bk/T \rb + \lb E_\bk/T \rb}{\lb E_\bk/T \rb^2} \,.
\end{equation}
which grows as $\sim e^{E_\bk/T}/{\lb E_\bk/T \rb^2}$ in the large-$E_\bk$ limit. A naive estimation using freeze-out temperature $T=160$ MeV and meson energy $1\text{ GeV}<E_\bk<2\text{ GeV}$ gives $7<A<860$, indicating that terms containing $A$ could have much larger contributions than the remaining parts. Such contributions, as will be shown in the next section, vanish at global equilibrium, but may play an important role when the fluid is at local equilibrium, especially for vector mesons at high momentum in realistic heavy-ion collisions.

\section{Global equilibrium and beyond}
\label{sec:check}

In order to check the obtained results and to better understand the physical contents of the 
formulae for vector and tensor spin polarization, it is useful to consider the global equilibrium
limit. Global equilibrium is notably independent of pseudo-gauge transformations and it is 
achieved by requiring the LEDO $\opt{\r}_\LE$ to be independent of the choice of hypersurface, 
which implies that the four-temperature vector fulfills the Killing equation
\[
 \pt_\mu \beta_\nu + \pt_\nu \beta_\mu = 0 \,,
\]
and the spin potential equals the thermal vorticity and is constant in space-time
\begin{equation*}
    \O_{\r\s}(x) = \varpi_{\r\s} \equiv  \frac{1}{2}\pt_{[\s}\b_{\r]}(x) 
    \,,\quad \pt_\m\O_{\r\s}(x)=0 \,.
\end{equation*}
The four-temperature is then given by
\begin{equation}\label{eq:thermal:current}
    \b_{\m}(x)=b_{\m} + \varpi_{\m\n}x^{\n}
\end{equation}
with $ b_{\m} $ and $ \varpi_{\m\n} $  being constants independent to $ x $\,.

Substituting \eq{eq:thermal:current} into \eqs{frsT2}{frsS2}, we obtain the first-order MVSD
\begin{equation}\label{effe1}
    f^{(1)}_{rs} (x,k) = -i\d(k^2-m^2)\h(k^0) n_B(1+n_B) \e^{\n}_{s}(k)\e^{\m*}_{r}(k) \varpi_{\m\n} \,.
\end{equation}
Now the thermal vorticity $ \varpi_{\m\n} $, as it is well known, can be decomposed into
two space-like vectors $ w^\m$ and $\alpha^\m$ by choosing a four-velocity vector $u$
\begin{equation}\label{thvdec}
    \varpi_{\m\n} = \e_{\m\n\r\s} w^{\r} u^{\s} +\alpha_\m u_\n - \alpha_\n u_\m \,.
\end{equation}
where
\[
    w^{\m}= -\frac{1}{2} \e_{\m \n\r\s} \varpi_{\n\r} u_\s \,, \qquad \qquad \alpha^\mu = \varpi^{\m\n}u_\n \,.
\]
If we choose as $u$ the particle four-velocity $k/m$, $w$, and $\alpha$ are the ``thermal 
angular velocity" and the ``thermal acceleration" four-vectors having only space components 
in the rest-frame of the particle. At the leading order in thermal
vorticity, from eqs.~\eqref{spinpolvec},~\eqref{eq:cooper-frye}, and \eqref{effe1}, we obtain
\begin{equation}\label{spin1stord}
  S^\m(k) \simeq -\frac{1}{3m}\e^{\m\n\r\s}\frac{\int_\Sigma d\Sigma \cdot k \; n_B (1+n_B) 
  k_\s\varpi_{\n\r}}{\int_\Sigma d\Sigma \cdot k \; n_B} = \frac{2}{3} 
  \frac{\int_\Sigma d\Sigma \cdot k \; n_B (1+n_B) w^\mu}{\int_\Sigma d\Sigma \cdot k \; n_B} \,.
\end{equation}
which is in agreement with the {\em ansatz} presented in ref.~\cite{Becattini:2016gvu}.

On the other hand, as it is well known, the leading term of the tensor polarization $\fT$ in \eq{exp:T1} at global equilibrium
is quadratic in thermal vorticity. Utilizing \eqs{final:S}{final:TT}, 
one can obtain the second-order MVSD at global equilibrium
\begin{equation}
\begin{split}\label{MVSD-second-global}
    f^{(2)}_{rs} (x,k) =& -\frac{1}{2}\d(k^2-m^2) \h(k^0) n_B(1+n_B)(1+2n_B) \e^{\n}_{s}(k)\e^{\m*}_{r}(k)\\
    & \times \ls \lb \w^{\r\s} -\frac{1}{m^2}k^{\r}k^{\s}\rb +\frac{1}{2m^2}k^{\r}k^{\s} \rs \varpi_{\r\m}
    \varpi_{\s\n} + \cdots \,,
\end{split} 
\end{equation}
where terms proportional to $ \d_{rs} $, which do not contribute to the polarization, have been omitted.
By using \eq{eq:spin-align} and \eq{frszero} for the zeroth order MVSD, one 
obtains the contribution to spin alignment in the phase space
\begin{equation}\label{dth00}
    \d\H_{00}(x,k) = -\frac{1}{6}(1+n_B)(1+2n_B)\ls\e^{\m*}_0(k)\e^{\n}_0(k) +\frac{1}{3} \Delta^{\m\n} \rs
    \ls \Delta^{\r\s} +\frac{1}{2m^2}k^{\r}k^{\s} \rs \varpi_{\r\m}\varpi_{\s\n} \,,
\end{equation}
where we have defined
\[
    \Delta^{\r\s} = \lb \w^{\r\s} -\frac{1}{m^2}k^{\r}k^{\s}\rb =  \lb \w^{\r\s} - u^\r u^\s \rb \,.
\]
Plugging the decomposition \eq{thvdec} with $ u = k/m $ into \eq{dth00}, one obtains
\begin{equation}\label{dth00-2}
    \d\H_{00}(x,k) = -\frac{1}{6}(1+n_B)(1+2n_B) \ls | \e_0 \cdot w |^2 + \frac{1}{3} w^2 + \frac{1}{2} |\e_0 \cdot \alpha |^2 + \frac{1}{6} \alpha^2 \rs \,.
\end{equation}
It can be shown that, at global equilibrium, the contribution from $\a$ in \eq{dth00-2} vanishes 
after integrating it over the freeze-out hypersurface, according to \eq{eq:cooperfrye}.
Indeed, both terms including thermal accelerations in \eq{dth00-2} give rise to integrals of the kind
\begin{align*}
    & \int_\S d\S\cdot k\; n_B(1+n_B)(1+2n_B) \alpha^\mu \alpha^\nu
    = \frac{1}{m^2}\int_\S d\S\cdot k\, n_B(1+n_B)(1+2n_B)k_{\r}k_{\s} 
    \varpi^{\r\m}\varpi^{s\n} \\
   &= \frac{1}{m^2}\int_\S d\S\cdot k\, \pt^{\m}_{x} \pt^{\n}_{x} n_B \,.
\end{align*}
Since at global equilibrium $ (k\cdot \pt_x) n_B = 0 $, the integrand is divergence-free and we
can transfer the integration over a hyperplane at constant Cartesian time $t$
\begin{equation}\label{nB:integral}
    \frac{1}{m^2}\int_\S d\S\cdot k\, \pt^{\m}_{x} \pt^{\n}_{x} n_B = \frac{k^0}{m^2}  
     \int_{x^0=t} d^3\bx ~~ \pt^{\m}_{x} \pt^{\n}_{x} n_B (k\cdot\b(x)) \,,
\end{equation}
provided that boundary terms vanish altogether. Again, provided that boundary terms vanish,
the right-hand side of \eq{nB:integral} vanishes for any choice of the indices except $ \m=\n=0 $.
However, in the latter case,
\[
   k^0 \int_{x^0=t} d^3\bx ~ \lb\frac{\pt}{\pt x^0}\rb^2 n_B = 
   \lb\frac{\pt}{\pt t}\rb^2 k^0 \int_{x^0=t} d^3\bx ~ n_B \,.
\]
Now $ k^0 \int_{x^0=t} d^3\bx\, n_B $ is a constant of time as $ (k\cdot \pt_x) n_B = 0 $ 
at global equilibrium, thus the above derivative vanishes. 
Therefore, we have the spin alignment at the global equilibrium
\begin{equation}\label{spin-globla-eq}
    \d\H_{00}(\bk) \approx -\frac{1}{6} \frac{\int d\S\cdot k\, f(x,k) (1+n_B)(1+2n_B)\ls | \e_0 \cdot w |^2 + \frac{1}{3} w^2 \rs}{\int d\S\cdot k\, f(x,k)} \,,
\end{equation}
which is in agreement with the exact form of the spin density matrix at the global 
equilibrium at all orders in thermal vorticity reported in ref.~\cite{Palermo:2023cup}.

For the general case at local equilibrium, other hydrodynamic fields, \eg ``shear-shear" terms or the gradient of the thermal shear and thermal vorticity, can also be the sources of the tensor polarization, as shown in the last section.
In order to isolate the contribution of each out-of-global-equilibrium source, we can turn on only that source and keep other out-of-global-equilibrium sources off.

As an example, let us consider the contribution of the gradient of the thermal vorticity, $ \pt_\a \varpi_{\m\n} $, in $f_{rs}^{(2)}$.
We substitute the hydrodynamic fields in \eqs{final:S}{MVSD:f2T} with
\[
    \pt_{\a_1} \O_{\r_1\s_1}\rightarrow \pt_{\a_1} \varpi_{\r_1\s_1},\quad \pt_{\a_1}\pt_{\a_2}\b_{\n_1} \rightarrow -\frac{1}{2}\pt_{(\a_1} \varpi_{\a_2)\n_1} \,,
\]
and obtain
\begin{equation}
\begin{split}
    f^{(2)}_{rs}{\big|}_{\pt\varpi}(x,k) =&~ \d(k^2-m^2) \h(k^0) n_B(1+n_B) \e^{r*}_{(\m}(k) \e^{s}_{\n)}(k)\\
    &~\times \frac{1}{4E_\bk} \htt^{[\r} [\pt^{\m]}\varpi_{\r\s}(x)] \lb 2\w^{\s\n} -\hat{k}^\s \htt^{\n}\rb  +\cdots \,,
\end{split}
\end{equation}
where we ignore the terms proportional to $ \d_{rs} $ as they do not contribute to the polarization at this order of gradients.
By using equation \eq{eq:spin-align}, we have the contribution to spin alignment in phase-space
\begin{equation}\label{spal-ptomega}
    \d\H_{00}{\big|}_{\pt\varpi}(x,k) = (1+n_B) \ls\e_{\m}^0(k)\e_{\n}^0(k) +\frac{1}{3} \D_{\m\n} \rs \frac{1}{6 E_\bk} \htt^{[\r} [\pt^{\m]}\varpi_{\r\s}(x)] \lb 2\w^{\s\n} -\hat{k}^\s \htt^{\n}\rb\,.
\end{equation}

Similarly, we consider the contribution of the gradient of thermal shear in spin alignment by substituting the hydrodynamics fields in \eq{MVSD:f2T} with
\[
    \pt_{\a_1}\pt_{\a_2} \b_{\n_1} \rightarrow \frac{1}{2}\pt_{(\a_1} \x_{\a_2)\n_1} \,,
\]
where $ \x $ is the thermal shear defined by $ \x_{\r\s} \equiv \pt_{(\r}\b_{\s)}/2 $ . We obtain the spin alignment in the phase space
\begin{equation}\label{spal-ptshear}
\begin{split}
    \d\H_{00}|_{\pt\x}(x,k) =&~ (1+n_B) \ls\e_{\m}^0(k)\e_{\n}^0(k) +\frac{1}{3} \D_{\m\n} \rs \frac{1}{6E_k} [\pt_{\a}\x_{\r\s}(x)]\\
    & \times \Big[ (2\htt^{\r} +\gamma_k^2\hat{k}^{\r}) \w^{\a\m}\w^{\s\n} +(\w^{\r(\s}\hat{k}^{\a)} +\gamma_k^2\hat{k}^{\a}\hat{k}^{\r}\hat{k}^{\s}) \htt^{\m}\htt^{\n}\\
    & -(\w^{\r(\a} +\htt^{\r}\hat{k}^{(\a} +\gamma_k^2\hat{k}^{\r}\hat{k}^{(\a}) \w^{\s)\m}\htt^{\n}\Big] \,.\\
\end{split}
\end{equation}
Then, we obtain the contribution of the gradient of net spin potential, the difference between spin potential and thermal vorticity, in spin alignment
by substituting the hydrodynamics fields in \eq{final:S} with
\[
    \O_{\r_1\s_1} \rightarrow \d\O_{\r_1\s_1} \equiv \O_{\r_1\s_1} -\varpi_{\r_1\s_1} \,.
\]
The spin alignment in the phase space is
\begin{equation}\label{spal-ptspin}
\begin{split}
    \d\H_{00}|_{\pt\d\O}(x,k) =&~ (1+n_B) \ls\e_{\m}^0(k)\e_{\n}^0(k) +\frac{1}{3} \D_{\m\n} \rs \frac{1}{6E_k}[\pt_{\a}\d\O\indices{_{\r}^{\n}}(x)] \\
    &~ \times \lb \hat{k}^{\a}\htt^{\r}{\htt}^{\m} -\g_k^2\,\hat{k}^{\r}\w^{\m\a} -\D^{\a\r}\htt^{\m}\rb \,.
\end{split}
\end{equation}

In this section, we have derived contributions from quadratic terms in thermal vorticity as well as gradients of thermal vorticity and thermal shear tensors. Inspired by the $\Lambda$ hyperons' global polarization \cite{STAR:2017ckg}, the thermal vorticity $\sim\mathcal{O}(10^{-2})$ in heavy-ion collisions, indicating that the spin alignment induced by the quadratic term in $\varpi$, given by Eq. (\ref{spin-globla-eq}) is $\sim 10^{-4}$, which is much smaller than the experiment observation \cite{STAR:2022fan}. However, contributions from Eqs. (\ref{spal-ptomega})-(\ref{spal-ptspin}) could be important when the thermal vorticity, thermal shear, or spin potential have significant spacetime variations around the freeze-out hypersurface, which is a situation that might be encountered in heavy-ion collisions. For example, ref.~\cite{Karpenko:2016jyx} illustrates that the $ \varpi_{tz} $ component contains a significant spatial variation at certain points on the freeze-out hypersurface. A rough estimation based on outcomes of ref.~\cite{Karpenko:2016jyx} indicates that $\partial_x \varpi_{tz}$ reaches a magnitude of approximately $ 0.3\,\mathrm{fm}^{-1} \approx 0.05\,\mathrm{GeV}$ over some part of the freezeout hypersurface. Assuming a typical energy scale of $ E_\bk\approx 1\,\mathrm{GeV} $, the ratio $\partial_x \varpi_{tz}/E_\bk\approx 0.05$ emerges, which is comparable to the thermal vorticity itself. Of course, the integration over the whole hypersurface may mitigate this correction, but this 
requires further quantitative investigation.

This estimation also suggests that the second-order hydrodynamic terms reported in \eqs{spal-ptomega}{spal-ptspin} 
may contribute substantially to spin alignment. Furthermore, the spin alignment induced by gradients of the thermal 
shear or net spin potential, given by  Eqs. (\ref{spal-ptshear}) and (\ref{spal-ptspin}), contains terms proportional 
to $\gk\equiv E_\bk/m$. For vector mesons with large momenta, these contributions would become more important, 
which also awaits verification from numerical simulations.

\section{Summary and outlook}
\label{sec:summary}

In summary, we have calculated the spin density matrix of neutral vector mesons up to 
the second order in the gradient expansion of the local thermodynamic equilibrium, \ie 
neglecting the dissipative part of the density operator. The tensor component of the spin 
density matrix, which is responsible for spin alignment, appears only at the second order in 
the gradient expansion, that is quadratic in the first-order derivatives or linear in the second 
order derivatives of the four-temperature field \( \beta_\mu \) and quadratic in the spin
potential \( \Omega_{\rho\sigma} \) or linear in its first-order derivative.
This calculation has been carried out in the so-called canonical pseudo-gauge, that is with 
the canonical stress-energy tensor and the canonical spin tensor coupled to a non-vanishing spin
potential $\Omega$.

To carry out our calculations, we have introduced the Matrix-Valued-Spin Distribution (MVSD) function 
defined from the Wigner function and developed a set of Feynman rules to compute them.
We found that, at this order of the expansion, the spin alignment parameter in the rest-frame of 
the particle reads
\begin{equation*}
    \H_{00}(x,k)-\frac{1}{3} \approx \frac{2f^{(2)}_{00}(x,k) - 2f^{(2)}_{11}(x,k)}{\sum_r f_{rr}^{(0)}(x,k)} \,,
\end{equation*}
where the zeroth-order MVSD $ f_{rs}^{(0)} $ is presented in \eq{frszero} and the second-order MVSD 
$ f^{(2)}_{rs} $ includes six different contributions
\begin{equation*}
    f^{(2)}_{rs} = f^{(2)}_{rs}|_{TT} +f^{(2)}_{rs}|_{TS} +f^{(2)}_{rs}|_{ST} 
    +f^{(2)}_{rs}|_{SS} +f^{(2)}_{rs}|_{T} +f^{(2)}_{rs}|_{S} \,.
\end{equation*}
The above sum represents contributions from different hydrodynamic fields and derivatives thereof 
and their final expressions are reported in \eqs{final:S}{final:TT}. We have also presented the full 
expression for the vector component of the spin density matrix - responsible for the mean spin 
polarization vector - and the tensor 
component of the spin density matrix. Additionally, we have discussed the vector and tensor spin polarization 
at global thermodynamic equilibrium and compared with previously obtained exact results, finding agreement 
at the second order in thermal vorticity expansion.

The formulae derived in this paper are in principle applicable to phenomenological calculations of vector 
meson spin alignment in relativistic heavy ion collisions to assess the contribution of second-order local 
equilibrium contributions. 

\appendix

\section{Spin density matrix and standard Lorentz transformation}
\label{sec:STDL}

In quantum field theory, the single-particle eigenvectors of four-momentum depend on the so-called {\em standard Lorentz transformation} (denoted by $[k]$) transforming the four-momentum $k_0$ with components $(m,{\bf 0})$ into 
the actual four-momentum $k$, \ie,
\[
    k = [k](k_0) \,. 
\]
One has \cite{Moussa:1966gjd}
\[
    \ket{k,r} = \widehat{[k]} \ket{k_0,r} \,,
\]
where $r$ is the spin index and, sometimes, this dependence is emphasized by writing the state as $\ket{[k],r}$ \cite{Moussa:1966gjd}.
By choosing a different standard Lorentz transformation $[k]'$\,, 
it can be readily shown that
\[
    \ket{[k]',r} = \sum_s D^S([k]^{-1}[k]')_{sr} \ket{[k],s}\,,
\]
where $[k]^{-1}[k]' \equiv R$ is a rotation (because $k_0$ is left unchanged by it) and $D^S$ its 
associated representation matrix of spin $S$. It is a straightforward consequence of the above
relation that creation operators also depend on the standard Lorentz transformation
\[
\widehat{a}_\bk^{r \dagger\prime} = \sum_s D^S(R)_{sr} \widehat{a}_\bk^{s \dagger }\,,
\]
and annihilation operators likewise. Therefore, the spin density matrix defined in
\eq{Spin-density-matrix} depends on the standard Lorentz transformation and it can be readily
shown that
\[
    \Theta^\prime(k)_{tu} = \sum_{rs} D^S(R^{-1})_{tr} \Theta(k)_{rs} D^S(R)_{su} \,,
\]
where the left-hand side is the spin density matrix defined with the primed creation and
annihilation operators according to \eq{Spin-density-matrix}.
The important question arises as to whether one can define a spin four-vector (and tensors as well
for $S \ge 1/2$), which is {\em objective}, that is independent of the standard Lorentz
transformation. Indeed, \eq{spinpolvec}, reporting the components of the mean spin vector
in the frame where the four-momentum of the particle is $k$, is independent 
of the rotation $R$
\begin{align*}
    S^{\prime\mu}(k) &\equiv \sum_{i=1}^3 \tr \left( D^S(J^i) \Theta^\prime(k) \right) [k]\indices{^{\prime\mu}_i} = 
    \sum_{i=1}^3 \tr \left( D^S(J^i) D^S(R^{-1}) \Theta(k) D^S(R) \right) [k]\indices{^{\prime\mu}_i}  \\
    &= \sum_{i=1}^3 \tr \left(D^S(R) D^S(J^i) D^S(R^{-1}) \Theta(k) \right) [k]\indices{^{\prime\mu}_i} =
    \sum_{i,j=1}^3 \tr \left(D^S(J^j)\Theta(k) \right) (R^{-1})\indices{^i_j} [k]\indices{^{\prime\mu}_i} \\
    &=\sum_{j=1}^3 \tr \left(D^S(J^j)\Theta(k) \right) [k]\indices{^{\mu}_j} = S^\mu(k)\,.
\end{align*}
In the above equalities, we have used the cyclicity of the trace and the well-known relation 
in group representation theory
\[
D^S(R) D^S(J^i) D^S(R^{-1}) = \sum_{j=1}^3 (R^{-1})\indices{^i_j} D^S(J^j)\,.
\]
Conversely, an expression like \eq{spinpolvec2} without further specification would not be 
independent of the standard Lorentz transformation
\begin{align*}
S^{i\prime}_{\rm rest}(k) &\equiv \tr \left( D^S(J^i) \Theta^\prime(k) \right) = 
\tr \left( D^S(J^i) D^S(R^{-1}) \Theta(k) D^S(R) \right)  \\
&= \tr \left(D^S(R) D^S(J^i) D^S(R^{-1}) \Theta(k) \right) =
\sum_{j=1}^3 \tr \left(D^S(J^j)\Theta(k) \right) (R^{-1})\indices{^i_j} \\
& \ne\tr \left(D^S(J^i)\Theta(k) \right) = S^{i}_{\rm rest}(k)\,.
\end{align*}
Similar arguments hold for all tensors defined from the spin density matrix, like the tensor 
polarization $\fT$ for vector bosons in \eq{exp:T1}.

\section{Derivative operator \texorpdfstring{$D$}{D}}
\label{sec:D}
In this section, we present some examples of how to calculate the Feynman diagrams.

Among the second-order diagrams, a typical derivative structure reads
\begin{equation}
\begin{split}
\label{D:diagram1}
\begin{tikzpicture}[baseline=-0.65ex]
    \node[TSnode]   at (0,0)            (T1)            [anchor=west,
                                                        label={100: $p_1$},
                                                        label={-100: $p_2$}]                    {$D^\a_{(1)}$};
    \node[TSnode]   at (1.8,0)          (T2)            [anchor=west,
                                                        label={90: $p_3$},
                                                        label={-90: $p_4$}]                     {};
    \node[fnode]    at (3.2,0)          (f1)            [anchor=west,
                                                        label={85: $p_0$},
                                                        label={-85: $p_5$}]                     {};
    \draw[->-]      (f1.north)      to [out=120,in=60]                                  (T1.north);
    \draw[->-]      (T1.south)      to [out=-60,in=120]                                 (T2.north);
    \draw[->-]      (T2.south)      to [out=-60,in=-120]                                (f1.south);
\end{tikzpicture}
\sim    & \int \prod_{i=0}^{5} d^3\bp_i\,\d(\bp_1-\bp_2) \lb-\frac{i}{2}\rb\lb \na^\a_{1} -\na^\a_{2}\rb \d(\bp_0-\bp_1)\\
        & \times \d(\bp_2-\bp_3)\d(\bp_3-\bp_4)\d(\bp_4-\bp_5)\d\lb \bk -\frac{\bp_0}{2} -\frac{\bp_5}{2}\rb \,.
\end{split}
\end{equation}
where all the $\delta$ represent 3-dimensional delta functions $\delta^{(3)}$, with superscripts ``$(3)$" being embedded for simplicity. Our purpose is to move the derivatives to the right-hand side (RHS) of the delta functions.
To this end, we first perform variable substitutions as
\begin{equation}
\begin{aligned}
    \bq_i &= \bp_i - \bp_{i+1},\, \mathrm{with}\,i=0-4 \,, \\
    \bq_5 &= \frac{\bp_0}{2} +\frac{\bp_5}{2} \,.
\end{aligned}
\end{equation}
We can express the substitutions in the matrix form,
\begin{equation}
    \bq_i = \sum_{j=0}^{5} A_{ij} \bp_j,\,\mathrm{with}\  A_{ij}=
    \begin{pmatrix}
        1 & -1 &  &  &  & \\
         & 1 & -1 &  &  & \\
         &  & 1 & -1 &  &\\
         &  &  & 1 & -1 & \\
         &  &  &  & 1 & -1\\
        \frac{1}{2} &  &  &  &  & \frac{1}{2}
    \end{pmatrix} \,.
\end{equation}
Such a matrix $ A $ is reversible ($ \det A =1$). Then the formula in \eq{D:diagram1} is converted to the following form,
\begin{align}
    \ref{D:diagram1} &= \int \prod_{i} d^3\bp_i\,\d (\bq_1)\, \lb \sum_{i=0}^{5}\na^\a_i v_i \rb\, \d (\bq_0) \d (\bq_2) \d (\bq_3) \d (\bq_4) \d (\bk-\bq_5) \non
    &= \int \prod_{i} d^3\bq_i\,\d (\bq_1)\, \lb \sum_{i,j=0}^{5}\na^\a_{(\bq),i} A_{ij} v_j \rb\, \d (\bq_0) \d (\bq_2) \d (\bq_3) \d (\bq_4) \d (\bk-\bq_5)
    \label{D:paAv}
\end{align}
with $ \bv=\lb-i/2\rb(0,1,-1,0,0,0)^{T} $, where we have used
\begin{equation}
    \na^\a_i = \sum_{j=0}^{5}\na^\a_{(\bq),j} A_{ji} \,,
\end{equation}
and the derivatives with respect to $ \bq_i $ is defined as
\begin{align}
    \na^\a_{(\bq),i} =\begin{cases}
        0,\,\mathrm{with}\ \a=0 \,,\\
        \frac{\pt}{\pt \bq_{i,\a}},\,\mathrm{else} \,.
    \end{cases}
\end{align}
In \eq{D:paAv}, derivatives with respect to $\bq_i$, $i=0,2,3,4,5$ will give boundary terms after integrating over $\bq_i$, which would in general be set to zeros. Only the $\na^\a_{(\bq),1}$ term survives in the calculation, 
\begin{equation}
    \ref{D:diagram1} = \int \prod_{i} d^3\bq_i\,\d (\bq_1)\, \lb \sum_{i=0}^{5}\na^\a_{(\bq),1} A_{1i} v_i \rb\,
    \d (\bq_0) \d (\bq_2) \d (\bq_3) \d (\bq_4) \d (\bk-\bq_5) \,.
\end{equation}
Then all the delta functions can be moved to the right-hand-side of $\na^\a_{(\bq),1}$
because $\delta (\bq_i)$, $i\neq 1$ is commutable with $\na^\a_{(\bq),1}$. 
Finally, we perform the variable substitution again and obtain
\begin{equation}
    \ref{D:diagram1} = \int \prod_{i}d\bp_i\, \d (\bq_0) \d (\bq_1) \d (\bq_2) \d (\bq_3) \d (\bq_4) \d (\bk-\bq_5)\, \lb \sum_{i,j,s,w=0}^{5}\na^\a_{i}A^{-1}_{ij} P_{js} A_{sw} v_w \rb \,.
\end{equation}
where $ P_{js}=\diag \{0,1,0,0,0,0\} $.
We then obtain 
\begin{equation}
    D^\a_{(1)} = \sum_{i,j,s,w=0}^{5}\na^\a_{i}A^{-1}_{ij} P_{js} A_{sw} v_w = \lb-\frac{i}{2}\rb \lb \na^\a_{0} +\na^\a_{1} -\na^\a_{2} -\na^\a_{3} -\na^\a_{4} -\na^\a_{5}\rb \,.
\end{equation}
This technique works even for multiple ``$D$"s because one can always deal with the ``$D$"s on the far right.

Several other examples are presented as follows.
\begin{equation}
\label{D:diagram2}
\begin{tikzpicture}[baseline=-0.65ex]
    \node[TSnode]   at (0,0)            (T1)            [anchor=west,
                                                        label={100: $p_1$},
                                                        label={-100: $p_2$}]                    {$D_{(1)}$};
    \node[TSnode]   at (1.8,0)          (T2)            [anchor=west,
                                                        label={90: $p_3$},
                                                        label={-90: $p_4$}]                     {};
    \node[fnode]    at (3.2,0)          (f1)            [anchor=west,
                                                        label={85: $p_0$},
                                                        label={-85: $p_5$}]                     {};
    \draw[->-]      (f1.north)      to [out=120,in=60]                                  (T1.north);
    \draw[-<-]      (T1.south)      to [out=-60,in=120]                                 (T2.north);
    \draw[->-]      (T2.south)      to [out=-60,in=-120]                                (f1.south);
\end{tikzpicture}
\quad D^\a_{(1)} = \lb-\frac{i}{2}\rb \lb \na^\a_0 +\na^\a_1 +\na^\a_2 +\na^\a_3 -\na^\a_4 -\na^\a_5 \rb \,.
\end{equation}
\begin{equation}
\label{D:diagram3}
\begin{split}
    \begin{tikzpicture}[baseline=-0.65ex]
    \node[TSnode]   at (0,0)            (T1)            [anchor=west,
                                                        label={100: $p_1$},
                                                        label={-100: $p_2$}]                    {};
    \node[TSnode]   at (1.8,0)          (T2)            [anchor=west,
                                                        label={90: $p_3$},
                                                        label={-90: $p_4$}]                     {};
    \node[TSnode]   at (3.2,0)          (T3)            [anchor=west,
                                                        label={90: $p_5$},
                                                        label={-90: $p_6$}]                     {$D_{(3)}$};
    \node[fnode]    at (4.6,0)          (f1)            [anchor=west,
                                                        label={85: $p_0$},
                                                        label={-85: $p_7$}]                     {};
    \draw[->-]      (f1.north)      to [out=120,in=60]                                  (T1.north);
    \draw[-<-]      (T1.south)      to [out=-60,in=120]                                 (T2.north);
    \draw[->-]      (T2.south)      to [out=-60,in=120]                                 (T3.north);
    \draw[->-]      (T3.south)      to [out=-60,in=-120]                                (f1.south);
\end{tikzpicture}& \\
\quad D_{(3)} = \lb-\frac{i}{2}\rb & \lb \na^\a_0 +\na^\a_1 -\na^\a_2 -\na^\a_3 +\na^\a_4 +\na^\a_5 -\na^\a_6 -\na^\a_7\rb \,.
\end{split}
\end{equation}

\section{Space-time reversal property of MVSD}
\label{sec:PTrev}
In this section, we will study the property of $f^{(n)}_{r,s}(x,k)$ under space-time reversal
(the combined transform of parity $\mathcal{P}$ and time reversal $\mathcal{T}$) as shown in \eq{eq:PTrev}.
This results from our assumption that for the integrals over the hypersurface, we only consider the contribution from $ \Xi $,
which is assumed to be a large enough flat space-like hypersurface.

Under spacetime reversal with respect to a certain point $ x $, the field operator and the derivative are transformed as
\begin{align}
\label{PT:A}
    \mathcal{PT}:\,&A_\m(x)\rightarrow A^{\dg}_\m(2x-y) \,,\\
\label{PT:partial}
    \mathcal{PT}:\,&\pt_\m \rightarrow -\pt_\m \,.
\end{align}
Since we consider neutral Proca particles, namely $ A=A^{\dg} $, the dagger is irrelevant for now, but it will be important later.
Based on \eqs{PT:A}{PT:partial}, the energy-momentum tensor and spin tensor are transformed as
\begin{align}
    \mathcal{PT}:\,&\, T^{\m\n}(y)\rightarrow T^{\m\n}(2x-y) \,,\\
    \mathcal{PT}:\,&\, S^{\m\r\s}(y)\rightarrow -S^{\m\r\s}(2x-y) \,.
\end{align}
Then the Wigner operator is transformed as follows
\begin{equation}
\begin{split}
    \cpt\, \opt{W}^{\m\n}_{+}(x,k) (\cpt)^{-1} =& \int d^4s\,e^{ik\cdot s} \opt{A}^{\m}(2x-y-\frac{s}{2}) \opt{A}^{\n}(2x -y +\frac{s}{2})\h(k^2)\h(k^0)\\
    =& \opt{W}^{\n\m}_{+}(2x-y,k) \,,
\end{split}
\end{equation}
from which we obtain the transformation properties of the MVSD operator
\begin{equation}
\begin{split}
    \cpt \opt{f}_{rs}(x,k) (\cpt)^{-1} &= \frac{1}{2\p}\e^{\m*}_{r}(k)\e^{\n}_{s}(k) \opt{W}^{\n\m}_{+}(2x-y,k)\\
    &= \frac{1}{2\p} (-1)^{r+s} \e^{\n*}_{-s}(k)\e^{\m}_{-r}(k) \opt{W}^{\n\m}_{+}(2x-y,k)\\
    &= (-1)^{r+s}\opt{f}_{-s,-r}(2x-y,k) \,.
\end{split}
\end{equation}

The $n$th-order MVSD is constructed by terms like 
\begin{equation}
\label{exmp:MVSD:2}
\begin{split}
    f^{(n)}_{rs} &\supset \ls\mathrm{Hydrodynamic\ parameters\ and\ intergrals\ over\ }\l\rs \ls\cdots\int d\Xi_{\m_l}(y_l)\cdots\rs\\
    & \times\ls\cdots(y_l-x)^{\a_m}\cdots\rs \lan \mathrm{Operators\ }(\opt{T}(y-i\l\b)\mathrm{\ or\ }\opt{S}(y-i\l\b))\opt{f}_{rs}(x,k)\ran_{0,c} \,.
\end{split}
\end{equation}
The sum of the number of $ (y-x) $ and the number of $ \opt{S} $ should be equal to $ n $.
After all, a $ (y-x) $ corresponds to a $ \pt $ on hydrodynamic parameters and an $ \opt{S} $ corresponds to a spin potential $ \O $.
For example, one can test these correspondences from \eq{def:B1} and \eq{def:B2}.

We construct the space-time reversal transformations of the operators and we choose the fixed point to be $ x $, the space-time coordinate of MVSD. We obtain
\begin{align}
    & \lan \opt{O}_1 \cdots \opt{O}_l \opt{f}_{rs}(x,k)\ran_{0,c} \non
    &= \lan (\cpt)^{-1}(\cpt)\opt{O}_1(\cpt)^{-1}(\cpt) \cdots (\cpt)^{-1}(\cpt) \opt{O}_l (\cpt)^{-1}(\cpt) \opt{f}_{rs}(x,k)\ran_{0,c} \non
    &= \lan (\cpt)\opt{O}_1(\cpt)^{-1} \cdots (\cpt) \opt{O}_l (\cpt)^{-1} (\cpt) \opt{f}_{rs}(x,k)(\cpt)^{-1}\ran_{0,c} \,,
    \label{PT:operator}
\end{align}
which is valid because $ \opt{\r}_0= \exp\{-\b\cdot\opt{P}\}/Z_0 $ is invariant under $ \mathcal{PT} $.
The operators are transformed according to
\begin{align}
    (\cpt)\opt{T}^{\m\n}(y-i\l\b)(\cpt)^{-1} =& \opt{T}^{\m\n\dg}(2x-y+i\l\b) = \opt{T}^{\m\n}(2x-y-i\l\b) \,, \non
    (\cpt)\opt{S}^{\m\r\s}(y-i\l\b)(\cpt)^{-1} =& -\opt{S}^{\m\r\s\dg}(2x-y+i\l\b) = -\opt{S}^{\m\r\s}(2x-y-i\l\b) \,.
\end{align}
Since we take $ \int_{\Xi} d\Xi_{\m}(y) $ on a flat space-like hyperplane, we can substitute the variables by $ y-x \rightarrow x-y $\,.
Then we observe that, in \eq{exmp:MVSD:2}, under space-time reversal,
a factor of $ (y-x) $ produces a minus sign, an operator $ \opt{S} $ also produces a minus sign, and the MVSD operator produces $ (-1)^{r+s} $\,.
Therefore, we have an extra factor of $ (-1)^{r+s+n} $\,, while the MVSD operator is transformed to $ \opt{f}_{-s-r}(x,k) $\,.
Finally, we conclude that
\begin{equation*}
    f^{(n)}_{r,s}(x,k) = (-1)^{r+s+n} f^{(n)}_{-s,-r}(x,k) \,.
\end{equation*}
%

\section{Contribution from time-like hypersurface}
\label{sec:time-like}
When deriving the Feynman rules, we only include the integrals contributed from $ \Xi $, the space-like part of $ \S_{\FO} $, which has been regarded as a large enough space-like hyperplane that is perpendicular to the time direction. 
In this section, we show that the diagram rules related to ``$ D $"s should be revised if one evaluates the Wigner function on the time-like part of the $ \S_{\FO} $.

In this section, we assume that the space-like hypersurface $ \Xi $ can be taken as a flat hyperplane $\Xi_A$ with small variations in time.
For any point on the auxiliary hyperplane $ \Xi_B $, the time component is a fixed value, 
\begin{equation}
    \htt\cdot x = t_B \,\mathrm{with}\, x\in \Xi_B \,,
\end{equation}
while the time for any point on $ \Xi_A$ is another fixed value,
\begin{equation}
    \htt\cdot x = t_B + \D t, \,\mathrm{with}\, x\in \Xi_A \,.
\end{equation}
Here $\Delta t$ is the distance between $\Xi_A$ and $\Xi_B$ on the time axis.
Meanwhile, for any point $ x $ on $ \S_{\FO} $\,, it should satisfy
\begin{equation}
    \htt\cdot x = t_B + \D t +\d t(x) \,,
\end{equation}
where $ \d t(x) $ is a small time deviation from the hyperplane $\Xi_A$\,.

We evaluate the integrals over space (\eg,~\eqs{example:D1}{example:D3}) again, by extending the integration domain from $ \Xi $ to $ \S_{\FO} $\,.
Using the Gauss's theorem, we decompose the integrals over $ \S_{\FO} $ as
\begin{equation}
\label{theorem:Gauss}
    \int_{\S_{\FO}} d\S_\m(y) = \int_{\Xi_{B}} d\Xi_\m(y) + \int_{\O}d\O(y) \frac{\pt}{\pt y^\m} \,,
\end{equation}
where the former integration is $ \htt_{\m} \int_{\Xi_{B}} d^3\bx $\,.
We assume that the integration over $ \O $ approximates to the integration over a square $ 4D $ region, whose space extent is infinitely large and whose boundaries in time direction are composed of $ \Xi $ and $ \Xi_A $\,.
Therefore, the integral over $\Omega$ approximates to
\begin{equation}
    \int_{\O}d\O(y) \frac{\pt}{\pt y^\m} \approx \int_{t_B}^{t_B+\D t}dy^0 \int d^3\by \frac{\pt}{\pt y^\m} \,.
\end{equation}
We calculate terms similar to~\eqs{example:D1}{example:D3}
\begin{equation}
    \int_{\S_{\FO}} d\S_{\m_l}(y_l)~ e^{-i(p_i-p_j)\cdot (y_l-x)}
    \sim \htt_{\m_l} (2\p)^3\d^{(3)}(\bp_i-\bp_j) \,,
\end{equation}
\begin{equation}
    \int_{\S_{\FO}} d\S_{\m_l}(y_l)~ e^{-i(p_i+p_j)\cdot (y_l-x)}
    \sim \htt_{\m_l} (2\p)^3\d^{(3)}(\bp_i+\bp_j) e^{i\htt\cdot(p_i+p_j)\d t} \,,
\end{equation}
\begin{equation}
\begin{split}
    & \int_{\S_{\FO}} d\S_{\m_l}(y_l)~ (y_l-x)_{\a_m} e^{-i(p_i-p_j)\cdot (y_l-x)} \\
    \sim~ &\htt_{\m_l}(2\p)^3\d^{(3)}(\bp_i-\bp_j) \lb-\frac{i}{2}\rb \ls \na_{i,\a_m} -\na_{j,\a_m} -2i\htt_{\a_m} \d t\rs e^{i\htt\cdot(p_i-p_j)\d t} \,,
\end{split}
\end{equation}
\begin{equation}
\begin{split}
    &\int_{\S_{\FO}} d\S_{\m_l}(y_l)~ (y_l-x)_{\a_m} e^{-i(p_i+p_j)\cdot (y_l-x)} \\
    \sim~& \htt_{\m_l} (2\p)^3\d^{(3)}(\bp_i+\bp_j) \lb-\frac{i}{2}\rb\ls\na_{i,\a_m} +\na_{j,\a_m} -2i\htt_{\a_m}\d t\rs e^{i\htt\cdot(p_i+p_j)\d t} \,,
\end{split}
\end{equation}
\begin{equation}
\begin{split}
    & \int_{\S_{\FO}} d\S_{\m_l}(y_l)~ (y_l-x)_{\a_m}(y_l-x)_{\a_n} e^{-i(p_i-p_j)\cdot (y_l-x)} \\
    \sim~& \htt_{\m_l} (2\p)^3\d^{(3)}(\bp_i-\bp_j) \lb-\frac{i}{2}\rb \ls \na_{i,\a_m} -\na_{j,\a_m} -2i\htt_{\a_m}\d t\rs \\
    & \times \lb-\frac{i}{2}\rb \ls \na_{i,\a_n} -\na_{j,\a_n} -2i\htt_{\a_m}\d t\rs e^{i\htt\cdot(p_i-p_j)\d t}
\end{split}
\end{equation}
with ``$ \sim $ " denoting that the calculation is done with the delta functions of the momenta and the integrals over the momenta. Compared with formulas in the main text, here we obtain additional terms from $\delta t$\,. Therefore, we conclude that, by extending the integration domain from $ \Xi $ to $ \S_{\FO} $~, the diagram rules should be revised.

We find that the new term from $\d t$ 
does not contribute to polarizations (including both vector and tensor) at $ \mathcal{O}(\pt) $~.
The vector polarization at $\mathcal{O}(\pt^2) $ will contain a new term proportional to $\d t$~,
while the tensor polarization is not affected by $\d t$ at $ \mathcal{O}(\pt^2) $~,
which is a consequence of the invariance of the tensor polarization under space-time reversal.

\acknowledgments

We thank Jian-Hua Gao, Shi Pu, Qun Wang, Xin-Qing Xie, and Shi-Zheng Yang for useful discussions.
Z.-H.Z. and X.-G.H. are supported by the Natural Science Foundation of Shanghai (Grant No. 23JC1400200), the National Natural Science Foundation of China (Grant No. 12147101, No. 12225502, and No. 12075061), and the National Key Research and Development Program of China (Grant No. 2022YFA1604900). F.B. and X.-L.S. are supported by the project PRIN2022
\textit{Advanced Probes of the Quark Gluon Plasma} funded by ''Ministero dell'Università e della Ricerca" and by ICSC – \textit{Centro Nazionale di Ricerca in High Performance Computing, Big 
Data and Quantum Computing}, funded by European Union – NextGenerationEU.

\paragraph{Note added.} Recently, we were informed of ref.~\cite{Yang:2024fkn}, which addresses a similar topic and was posted on arXiv on the same day.
According to the authors of it, our \eqs{final:SS}{final:TT} are consistent with their eq. (81), while, our eqs.~\eqref{final:S} and \eqref{MVSD:f2T} are not considered in ref.~\cite{Yang:2024fkn}.


\bibliographystyle{JHEP}
\bibliography{biblio.bib}

\end{document}